\documentclass[10pt,journal,compsoc]{IEEEtran}

\ifCLASSOPTIONcompsoc
  \usepackage[nocompress]{cite}
\else
  \usepackage{cite}
\fi
\ifCLASSINFOpdf
\else
\fi
\usepackage{csquotes}
\usepackage{amsmath,amsfonts}
\usepackage{array}
\usepackage[caption=false,font=normalsize,labelfont=sf,textfont=sf]{subfig}
\usepackage{textcomp,amsmath,amssymb}
\usepackage{stfloats}
\usepackage{url}
\usepackage{verbatim}
\usepackage{graphicx}
\usepackage{cite}
\usepackage{multirow}
\usepackage{balance}
\usepackage{booktabs,caption}
\usepackage[flushleft]{threeparttable}
\usepackage[framemethod=tikz]{mdframed}
\mdfdefinestyle{mpdframe}{
    frametitlebackgroundcolor   =black,
    frametitlerule              =true,
    roundcorner                 =5pt,
    middlelinewidth             =1pt,
    innermargin                 =0.3cm,
    outermargin                 =0.3cm,
    innerleftmargin             =0.3cm,
    innerrightmargin            =0.3cm,
    innertopmargin              =0.2cm,
    innerbottommargin           =0.2cm
}
\newcommand{\subsubsubsection}[1]{\paragraph{#1}}
\newcommand{\code}[1]{\textbf{#1}}
\begin{document}
%
\title{Sustainability Competencies and Skills in Software Engineering: An Industry Perspective}

%
%
%
%

\author{Rogardt Heldal, Ngoc-Thanh Nguyen, Ana Moreira, Patricia Lago, Leticia Duboc, Stefanie Betz, Vlad C. Coroamă, Birgit Penzenstadler, Jari Porras, Rafael Capilla, Ian Brooks, Shola Oyedeji, Colin C. Venters
\IEEEcompsocitemizethanks{
\IEEEcompsocthanksitem 
Rogardt Heldal and Ngoc-Thanh Nguyen are with the Western Norway University of Applied Sciences, Norway. Email: Rogardt.Heldal@hvl.no and Ngoc.Thanh.Nguyen@hvl.no.
\IEEEcompsocthanksitem Ana Moreira is with NOVA University Lisbon, Portugal. Email: amm@fct.unl.pt.
\IEEEcompsocthanksitem Patricia Lago is with Vrije Universiteit Amsterdam, Netherlands. Email: p.lago@vu.nl.
\IEEEcompsocthanksitem Leticia Duboc is with La Salle BCN, Ramon Llull University, Spain. Email: lduboc@salleurl.edu.
\IEEEcompsocthanksitem Stefanie Betz is with Furtwangen University, Germany. Email: besi@hs-furtwangen.de.
\IEEEcompsocthanksitem Vlad C. Coroamă is with Technische Universität Berlin, Germany. Email: coroama@tu-berlin.de.
\IEEEcompsocthanksitem Birgit Penzenstadler is with Chalmers University of Technology, Sweden. Email: birgitp@chalmers.se.
\IEEEcompsocthanksitem Jari Porras and Shola Oyedeji are with LUT University, Finland. Email: jari.porras@lut.fi and shola.oyedeji@lut.fi.
\IEEEcompsocthanksitem Rafael Capilla is with Rey Juan Carlos University, Spain. Email: rafael.capilla@urjc.es
\IEEEcompsocthanksitem Ian Brooks is with the University of the West of England Bristol, U.K. Email: ian.brooks@uwe.ac.uk.
\IEEEcompsocthanksitem Colin C. Venters is with the University of Huddersfield, U.K. and the European Organization for Nuclear Research (CERN), Switzerland. Email: c.venters@hud.ac.uk; c.venters@cern.ch.
\IEEEcompsocthanksitem Corresponding author: Rogardt Heldal (Rogardt.Heldal@hvl.no)
}
}

\IEEEtitleabstractindextext{%
\begin{abstract}
\textbf{Context:} Achieving the UN Sustainable Development Goals (SDGs) demands a shift by industry, governments, society, and individuals to reach adequate levels of awareness and actions to address sustainability challenges. Software systems will play an important role in moving towards these targets. Sustainability skills are necessary to support the development of software systems and to provide sustainable IT-supported services for citizens. 
\textbf{Gap:} While there is a growing number of academic bodies, including sustainability education in engineering and computer science curricula, there is not yet comprehensive research on the competencies and skills required by IT professionals to develop such systems.
\textbf{Research goal:} This study aims to identify the industrial sustainability needs for education and training from software engineers’ perspective. For this we answer the following questions: (1) what are the interests of organisations with an IT division with respect to sustainability? (2) what do organisations want to achieve with respect to sustainability, and how? and (3) what are the sustainability-related competencies and skills that organisations need to achieve their sustainability goals? 
\textbf{Methodology:} We conducted a qualitative study with interviews and focus groups with experts from twenty-eight organisations with an IT division from nine countries to understand their interests, goals and achievements related to sustainability, and the skills and competencies needed to achieve their goals. 
\textbf{Results:}  Our findings show that organisations are interested in sustainability, both idealistically and increasingly for core business reasons. They seek to improve the sustainability of processes and products but encounter difficulties, like the trade-off between short-term financial profitability and long-term sustainability goals or an unclear understanding of sustainability concepts. To fill these gaps, they have promoted in-house training courses, collaborated with universities, and sent employees to external training. The acquired competencies should support translating environmental and social benefits into economic ones and make sustainability an integral part of software development. We conclude that educational programs should include knowledge and skills on core sustainability concepts, system thinking, soft skills, technical sustainability, building the business case for sustainability, sustainability impact and measurements, values and ethics, standards and legal aspects, and advocacy and lobbying. 


\end{abstract}



}

\maketitle

\IEEEdisplaynontitleabstractindextext

%
\IEEEpeerreviewmaketitle

\ifCLASSOPTIONcompsoc
\IEEEraisesectionheading{\section{Introduction}\label{sec:introduction}}
\else
\section{Introduction}
\label{sec:introduction}
\fi

Digitalisation is pervasive and can either help or hinder the United Nations Sustainable Development Goals (SDGs)\footnote{\url{https://sdgs.un.org/goals}}\cite{seele2017game, coroama:2019:digital-rebound}. 
Organisations understand that but struggle in implementing sustainability in their service portfolio and their business practices~\cite{escoto2022refocusing,bocken2020barriers}. Consequently, there is a need to understand which competencies and skills industry requires, and how they can be integrated into their practices.
These new competencies and skills must be acquired through adequate learning programmes and courses addressing the different sustainability dimensions, i.e. environmental, economic, social, technical, and individual~\cite{becker2015sustainability}. For software engineers\footnote{The term \enquote{software engineer} in this article includes anyone who takes part in the process of designing, producing, and managing software.}, this ranges from the more technical aspects supporting Green IT and software sustainability to more social and individual ones facilitating software-driven processes in society.

Academia has made efforts to introduce sustainability in regular computer science programmes, as well as suggesting the skills and competencies needed by their students~\cite{watson2013assessing,stone2019sustainability,rogers2015using}.
According to these studies, future software engineers need to develop a sustainability mindset and acquire sustainability competencies able to produce sustainable IT-based systems or systems that both support more sustainable processes and monitor the achieved sustainability goals~\cite{mann2011green}.
However, industry is still unclear on which sustainability skills different sectors require to achieve their sustainability goals. 

Recent 
non-academic literature highlights the role and importance of skills for sustainability. For instance, even the British Plastic Federation~\cite{Holcroft2022the} mentions that the sustainability skills of employees are key for any strategy oriented to achieving a sustainable business. Similarly, Terrafiniti~\cite{terrafinity}, a sustainability consultancy, highlights that effective sustainability performance demands sustainability skills and competencies --- not only from sustainability professionals but also in other roles within the organisation. Hence, we not only need to identify which skills are more relevant in delivering sustainability in a particular organisational unit but also related units must recognise sustainability 
as a goal of the company's core business.

What prompts this study is that we, the authors, are under the impression that across industry there is (1) only a partial understanding of sustainability and there is (2) a limited understanding of how to address the lack of related competencies.
Additionally, across academia, there is (3) a lack of understanding of the needs of industry related to sustainability and (4) a need for a concrete teaching curriculum that could lead to the high-quality sustainability education which software engineers require.

This work aims to investigate the industrial sustainability needs for education and training from a software engineering perspective. To achieve this, we addressed the following three research questions: RQ1: \textit{What are the interests of organisations with an IT division with respect to sustainability?}; RQ2: \textit{What do organisations want to achieve with respect to sustainability, and how?}; and RQ3:\textit{ What are the sustainability-related competencies and skills that organisations need to achieve the established sustainability goals?} 
To this end, we interviewed sustainability and IT experts from twenty-eight (28) organisations in nine (9) different countries. 
Our main contributions are:

\begin{itemize}
\item A far-reaching overview of the organisations' perspective on sustainability, including (i) their general interest in sustainability; (ii) the sustainability goals they want to achieve; (iii) their achievements towards these goals and the difficulties faced in achieving them; (iv) the sustainability skills and competencies they already possess in-house and those that are missing; and (v) solutions to acquire the missing.
\item Initial insights on the gaps in current academic and non-academic training programmes for software engineers, and our recommendations to address those gaps for those who design the new programmes to enable future software engineers to achieve sustainability skills and competencies.
\end{itemize}

The rest of the paper is structured as follows: Section~\ref{sec:background} provides a comprehensive background of the concept of sustainability. Section~\ref{sec:research_method} elaborates on the employed research method. Section~\ref{sec:results} presents the results regarding competencies and skills. Section~\ref{sec:interpretation} interprets the study's findings. Section~\ref{sec:discussion} offers recommendations for training programs to address identified gaps in competencies and skills. Section~\ref{sec:threats} provides an analysis of the threats to validity. Section~\ref{sec:relatedWork}
offers a review of related work. Lastly, Section~\ref{sec:conclusion} concludes the study and highlights potential future research directions.


\section{Background}\label{sec:background}
This section starts with some background on the general notion of sustainability and follows with specific overviews of sustainability in IT and then Software Engineering. 

Although the principles of sustainability have been known to numerous human cultures throughout history, their first scientific usage was most likely in H.C. von Carlowitz's principles of sustainable forestry from 1713~\cite{von1732sylvicultura} (summarised in \cite{morgenstern2007:on-carlowitz}). 
As Hilty and Aebischer~\cite{hilty2015ict} comment, as the understanding at the time was that forests have one purpose, to produce wood, Carlowitz's basic principle is quite straightforward: \enquote{\textit{do not cut more wood than will grow in the same period of time}}. Of course, we know today that a forest accomplishes many further functions (such as producing oxygen, filtering air and water, preserving biodiversity, recreational and aesthetic values, and many more), which makes the sustainability perspective much more complex. The paradigm, however, is unchanged: As Venters et al.~\cite{venters2018software} discuss, the verb \enquote{to sustain} and the noun \enquote{sustainability} come from the Latin \enquote{sustenere}, which was used for both \enquote{to endure} and \enquote{to uphold} something. Hence, \enquote{sustainability} refers to the capacity of a system to endure for a certain amount of time~\cite{venters2018software}. Within the conceptualisation of sustainability put forward by the Brundtland Commission in 1987~\cite{Brundtland1987}, the system in question is Earth itself and the period of time, while not exactly specified, includes many generations into the future. The Brundtland definition thus encompasses two aspects: distributive justice (\enquote{\textit{the essential needs of the world's poor, to which overriding priority must be given}}~\cite{Brundtland1987}), but also intergenerational justice, for which the preservation of the biosphere is a prerequisite.

The relationship between the IT sector (or digitalisation in general) and sustainability has been conceptualised in various ways and under different names. Early concerns with the environmental footprint of the IT sector itself are usually referred to as \enquote{Green IT}, while the purposeful deployment of IT to reduce the environmental footprint in other economic or societal sectors is often called ``Green by IT''~\cite{coroama2009energy}. Other terms used to describe the latter are, for example, ICT4EE (ICT for energy efficiency), \enquote{Energy Informatics}~\cite{hilty2015ict} or \enquote{I(C)T enabling effect}~\cite{malmodin2014considerations}. Numerous further names describe the relationship between digitalisation and sustainability in general, which includes both the concepts of \enquote{Green IT} and \enquote{Green by IT}, but also the further dimensions of sustainability, in particular, the social one. Such names include \enquote{Digital Sustainability}, \enquote{Sustainable Computing} or \enquote{ICT for Sustainability (ICT4S)}~\cite{hilty2015ict}.

For the \enquote{software and sustainability} domain, there are also two views, which are quite similar to those of the broader \enquote{IT and sustainability} field~\cite{lago2013exploring}: one looking at the sustainability of software itself (foremost, thus, a technical notion of sustainable software), the other at deploying software engineering for sustainability (SE4S) beyond the software systems themselves~\cite{venters2018software}. Acknowledging both views, the \enquote{Karlskrona Manifesto for Sustainability Design} extends the well-known three dimensions of sustainability by another two: technical (to account for the desired long-term use of software) and individual (addressing personal freedom, dignity and fulfillment) for a total of five dimensions~\cite{becker2015sustainability}. While the individual dimension is not always represented, most literature in the field accounts for both the technical as well as the three established dimensions (environmental, economic, and social)~\cite{lago2015framing}. As is the case in general with sustainability, the dimensions are not entirely independent and there are often trade-offs among them~\cite{becker2015requirements}. And while current software engineering practice gives high value to the technical and economic dimensions, the social and environmental ones (and thus the crucial components of the sustainability concept as understood by the Brundtland commission) are often ignored\cite{lago2015framing}.


\section{Study Design and Research Method}\label{sec:research_method}
\begin{figure*}[h]
  \centering
  \includegraphics[width=.9\linewidth]{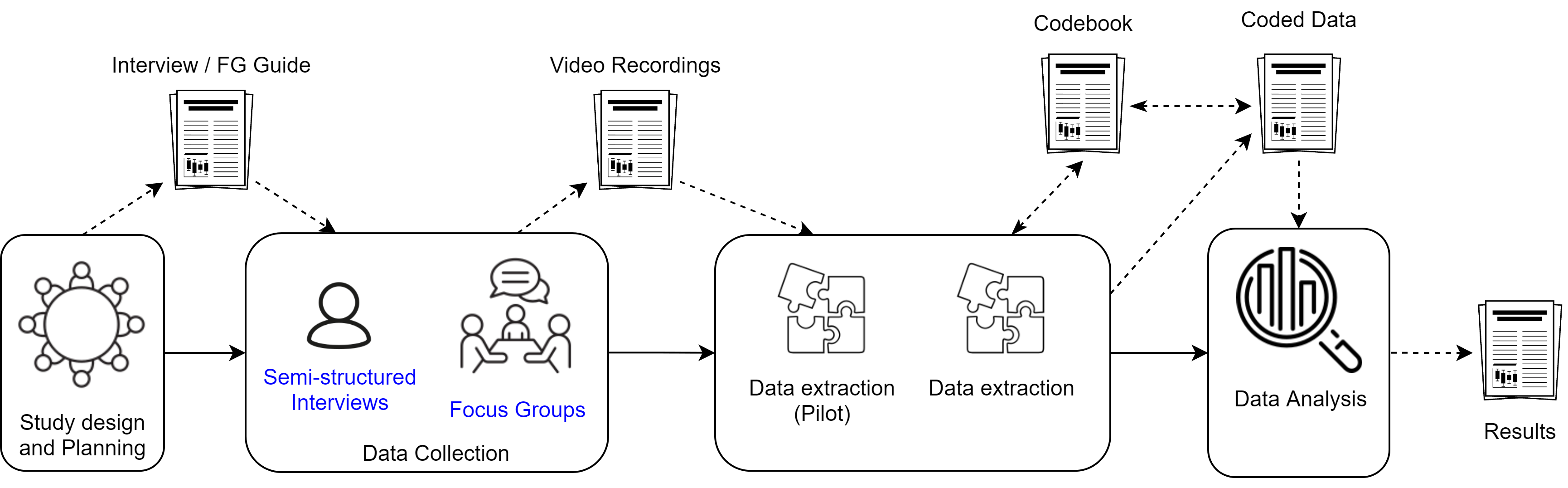}
  \caption{Study design and execution}
  \label{fig:study_design}
\end{figure*}

To answer our research questions, we used a mixed-methods approach~\cite{easterbrook2008selecting}, combining individual interviews and focus group interviews in a semi-structured format. For the sake of brevity, both individual interviews and focus group interviews are referred to as interviews hereafter. Our study process is illustrated in Figure~\ref{fig:study_design}. In summary, we extensively discussed our research goals and steps (study design and planning), creating a set of PowerPoint slides to guide our conversations in all interviews and focus groups (data collection). 
Additionally, we did a pilot study that provided a baseline structure for the subsequent interviews. 
All were recorded and transcribed and the relevant information was retrieved with the support of a code book (data extraction). Finally, the coded data was analysed and the results are presented.

\subsection{Goals and Research Questions}\label{sec:research_method:RQs}
To address our eventual goal (i.e., design education programmes that teach the required sustainability competencies and skills for future software engineering), we first need to understand what are the needs of the field, i.e., from industry\footnote{In general, with the term \enquote{industry} we mean practice from both the private and public sectors.}. Accordingly, we formulate the following overarching research question (RQ): \enquote{\textit{What are the industrial sustainability needs for education and training from software engineers' perspective?}}.

We break down RQ into three research sub-questions which guide our data collection:

RQ1: \textit{What are the interests of organisations with an IT division  with respect to sustainability?} The sustainability focus depends on the specific business domain and priorities. In this respect, the sustainability perspective depends on their specific interests and stakes. RQ1 helps us define the possible scope of future education programmes.

RQ2: \textit{What do organisations want to achieve with respect to sustainability, and how?} Sustainability can add significant value to both private and public organisations. However, to achieve this aim, sustainability must be tailored and embedded in the DNA of the organisation itself, for example, its business goals, values, and vision of the future market. Accordingly, this research question investigates the target achievements (what the organisations aim to achieve with respect to sustainability), the 
influence of software/ICT on these achievements, as well as the difficulties they face and expect. RQ2 helps us define and prioritise the various foci of future education programmes (e.g., creation of innovation, acquisition of new markets, compliance with regulation).

RQ3: \textit{What are the sustainability-related competencies and skills that organisations need to achieve the established sustainability goals?} To different extents, organisations are becoming aware of the sustainability-related competencies and skills that they have already in-house or that they miss in order to achieve their goals. This research question investigates the gaps in the IT workforce and, if applicable, the strategy organisations have in place or envisage to acquire the missing competencies and skills. RQ3 helps us define future education programmes' types and contents (e.g., mono- versus interdisciplinary, higher education versus professional training).

\begin{figure}[h]
  \centering
  \includegraphics[width=\linewidth]{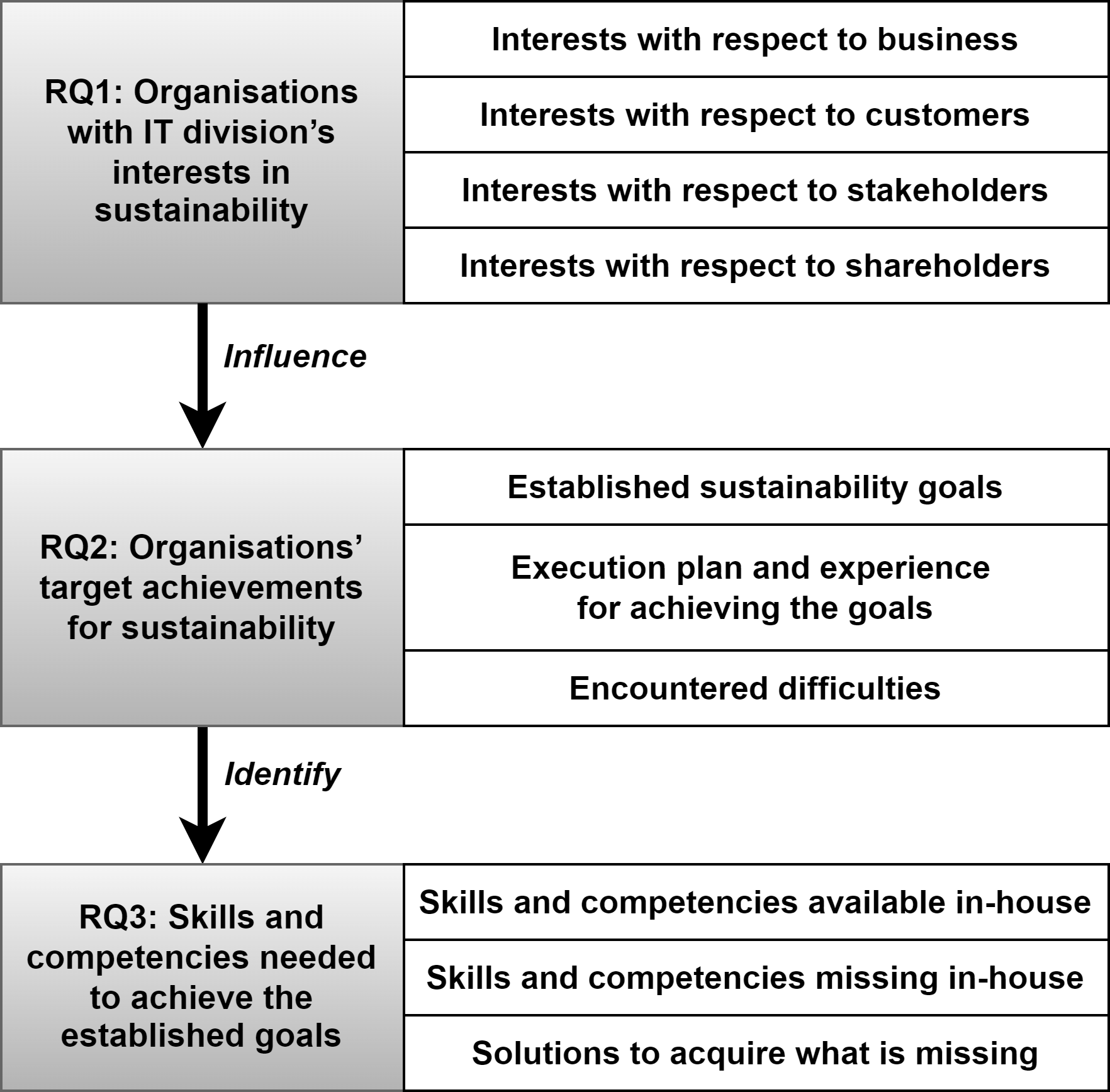}
  \caption{Themes with respect to research questions}
  \label{fig:RQ_relationship}
\end{figure}

Figure~\ref{fig:RQ_relationship} shows the relationship between RQs and the themes derived from the interviews. In Section~\ref{sec:results}, we will report in detail the findings related to each theme. 

\subsection{Data collection and analysis}\label{sec:research_method:mixed-method}
\subsubsection{Data collection}
To collect data, we conducted nine individual interviews and seven focus group interviews in a semi-structured format with industry practitioners. We contacted and recruited the participants by using our professional contacts. Individual interviews were employed due to the familiarity of researchers and interviewees available in their networks.  We 
conducted several focus group interviews to catalyse discussions among interviewees. Our selected ICT organisations have supported or participated in sustainability initiatives or have an ICT department involved in sustainability actions as part of the strategy of the company. We selected organisations from different countries and domains, as listed in Table~\ref{tab:interviewees}, to diversify the perspectives regarding sustainability. The organisations are anonymised to maintain confidentiality. In total, we interviewed 28 experienced IT/sustainability practitioners from 28 distinct organisations in different industrial domains belonging to 9 countries. We followed the statistical classification of economic activities in the European Community~\cite{rev20162} to classify the business sector of the organisations. To classify the organisation sizes, we followed the OECD scheme proposed in~\cite{entrepreneur}. Our participating organisations cover a wide spectrum of areas from software to telecommunications and resource supply. While most of them are private, nearly a third of the organisations (9/28) are from the public sector.


{\small 
\begin{table}[h]
  \caption{Organisations (anonymised) interviewed per country}
  \label{tab:interviewees}
  \begin{tabular}{p{0.2cm}p{1.3cm}p{3.5cm}p{1cm}l}
    \toprule
    \textbf{ID} &  \textbf{Country} & \textbf{Sector} & \textbf{Type} & \textbf{Size} \\
    \midrule
1 	&	Colombia 	&  Technology consultancy & Private &  $<$50\\ 
2 	&	Finland 	&  Software consultancy & Private & $<$250\\ 
3 	&	Finland 	&  Software & Private & $<$50\\ 
4 	&	Finland 	&  Software consultancy & Private & $<$250\\ 
5 	&	Germany 	&  Technology & Public & $<$50\\ 
6 	&	Germany 	&  Technology & Private & $<$50\\ 
7 	&	Germany 	&  Technology & Private & $<$50\\ 
8 	&	Netherlands &  Software consultancy & Private & $<$250\\ 
9 	&	Netherlands &  Public administration and defense & Public & 250+\\ 
10 	&	Netherlands &  Software consultancy & Private & 250+\\ 
11 	&	Norway 		&  ICT industry representative & Public & 250+\\ 
12 	&	Norway 		&  Energy provider & Public & 250+\\ 
13 	&	Norway 		&  Mobility provider & Public & 250+\\ 
14 	&	Norway 		&  Software consultancy & Private & 250+\\ 
15 	&	Norway 		&  Software consultancy & Private & $<$50\\ 
16 	&	Norway 		&  Waste management & Public & 250+\\ 
17 	&	Norway 		&  Technology & Public & 250+\\ 
18 	&	Portugal 	&  Software & Public & $<$250\\ 
19 	&	Portugal 	&  Software and Technology & Private & $<$250\\ 
20 	&	Portugal 	&  Software and Technology & Private & $<$50\\ 
21 	&	Portugal 	&  Software consultancy & Private & $<$50\\ 
22 	&	Spain 		&  Water supplier & Public & 250+  \\ 
23 	&	Spain 		&  Marketing and Advertising & Private & $<$250\\ 
24 	&	Spain 		&  Mobility provider & Private & 250+\\ 
25 	&	Sweden 		&  Networking and Telecommunication & Private & 250+\\ 
26 	&	Sweden 		&  Telecommunication & Private & 250+\\ 
27 	&	UK 			&  Technology & Private & 250+\\ 
28 	&	UK 			&  Technology & Private & $<$50\\  
    \bottomrule
\end{tabular}
\end{table}
}

Our participants 
have significant industry experience and have different roles and positions in their organisations, as shown in Table~\ref{tab:interview_companies}. The second column shows the business model with respect to the sustainability of their organisations, which is elaborated in more detail in Section~\ref{sec:results:demographics:model}. 

The majority of the participants are seniors with more than ten years of professional experience, and many have a computer science background or degree. We used online teleconferencing tools (e.g. Microsoft Teams, Zoom, Skype) to interview the participants. At the beginning of the interview, we took around five minutes to explain the goals of the interview. The prepared interview 
questions\footnote{Interview guide access link: \url{https://bit.ly/390MQju}} were then asked one by one.
The interviews were conducted from March to September 2021 and recorded with the consent of the interviewees. Individual interviews lasted between 30 minutes to 2 hours, while focus group interviews took a bit longer time as more discussion arose. The recorded interviews were transcribed manually or automatically by using, for example, Microsoft Office 365 (i.e., tool), depending on the researchers' preference. For automatic transcriptions, the responsible researchers spent time manually correcting transcription mistakes in the tool to ensure the quality of the research.

{\small 
\begin{table*}[h]
  \centering
  \caption{List of participants per company}
  \label{tab:interview_companies}
  \begin{tabular}{p{1.5cm}p{3.5cm}p{4cm}p{1cm}p{1cm}p{1.5cm}p{1.5cm}}
    \toprule
    \textbf{Organisation (Org.) ID} & \textbf{Business model with \newline respect to sustainability} & \textbf{Participant's role} & \textbf{Gender} & \textbf{Age} & \textbf{Years of \newline experience} & \textbf{Years in the company} \\
    \midrule
1 	& Producer & CEO and Consultant 			& Male & [40-50] & [20-30] & [0-10] \\ 
2 	& Producer & Account manager 				& Male & [40-50] & [10-20] & [0-10] \\ 
3 	& Producer & Senior advisor (ex. CEO) 		& Female & [40-50] & [10-20] & [0-10] \\ 
4 	& Producer & Sustainability manager 		& Male & [50-60] & [20-30] & [0-10] \\ 
5 	& Producer & Principal researcher 			& Male & [20-30] & [0-10] & [0-10] \\ 
6 	& Producer \& Consumer     & CEO 							& Male & [40-50] & [10-20] & [0-10] \\ 
7 	& Consumer & CTO							& Male & [30-40] & [10-20] & [0-10] \\ 
8 	& Producer \& Consumer     & CEO and Solution architect 	& Male & [40-50] & [20-30] & [10-20] \\ 
9 	& Consumer & Program manager 				& Male & [50-60] & [20-30] & [0-10] \\ 
10 	& Producer \& Consumer    & Enterprise architect 			& Male & [50-60] & [30-40] & [20-30] \\ 
11 	& Producer \& Consumer    & Director 						& Female & [30-40] & [10-20]  & [0-10]  \\ 
12 	& Producer \& Consumer    & Head of Strategy 				& Male & [40-50] & [20-30] & [0-10] \\ 
13 	& Producer \& Consumer    & Head of IT department 		& Male & [30-40] & [10-20] & [10-20] \\ 
14 	& Producer & Consultant 					& Male & [30-40] & [20-30] & [0-10] \\ 
15 	& Producer \& Consumer    & Director and Consultant 		& Female & [50-60] & [10-20] & [10-20] \\ 
16 	& Producer \& Consumer    & Chief data officer 			& Male & [40-50] & [20-30] & [0-10] \\ 
17 	& Producer \& Consumer    & Enterprise architect 			& Male & [50-60] & [30-40] & [0-10] \\ 
18 	& Producer & CEO 							& Male & [50-60] & [20-30] & [20-30] \\ 
19 	& Producer & CEO        					& Male & [50-60] & [30-40] & [5-10] \\ 
20 	& Producer & CEO       					& Male & [30-40] & [10-20] & [5-10] \\ 
21 	& Producer & CTO        					& Male & [50-60] & [20-30] & [5-10] \\ 
22 	& Producer & Environment  division chief 	& Male & [50-60] & [20-30] & [10-20] \\ 
23 	& Consumer & Sustainability manager 		& Female & [40-50] & [20-30] & [10-20] \\ 
24 	& Producer & Sustainability manager 		& Female & [30-40] & [10-20] & [0-10] \\ 
25 	& Producer & Principal researcher 			& Female & [40-50] & [20-30] & [20-30] \\ 
26 	& Producer & Environmental manager 		& Male & [50-60] & [30-40] & [20-30] \\ 
27 	& Producer & Head of External Collaborations & Male & [40-50] & [20-30] & [0-10] \\ 
28 	& Producer & CTO 							& Male & [40-50] & [20-30] & [0-10] \\
    \bottomrule
\end{tabular}
\end{table*}
}

\subsubsection{Data extraction and analysis}

To analyse the interviews, we employed the thematic data analysis approach proposed in~\cite{vaismoradi2016theme}. To facilitate the data analysis, we utilised Saturate App\footnote{\url{http://www.saturateapp.com/}}, which is a web-based platform that supports collaborative qualitative analysis
. It allows many researchers simultaneously to perform activities related to coding, labelling, and categorisation. The data analysis process was carried out as follows: Firstly, the transcripts were imported to Saturate App. Then, one researcher created the first version of a codebook in an Excel spreadsheet based on the interview questions. During the \textit{data extraction pilot} stage, 
we performed initial coding for the interview-based data collected from their interviews. They also validated and extended the codebook as needed until it was deemed stable by all coders. Finally, during the \textit{data extraction} stage, ten researchers involved in this study were divided into three sub-groups, each having at least three members. Each sub-group analysed one research question defined in Section~\ref{sec:research_method:RQs}. The groups conducted several workshops to validate and refine the coding done in the first stage so that the original coding for all interviews was verified and agreed on by several researchers. 

\begin{figure}[h]
  \centering
  \includegraphics[width=\linewidth]{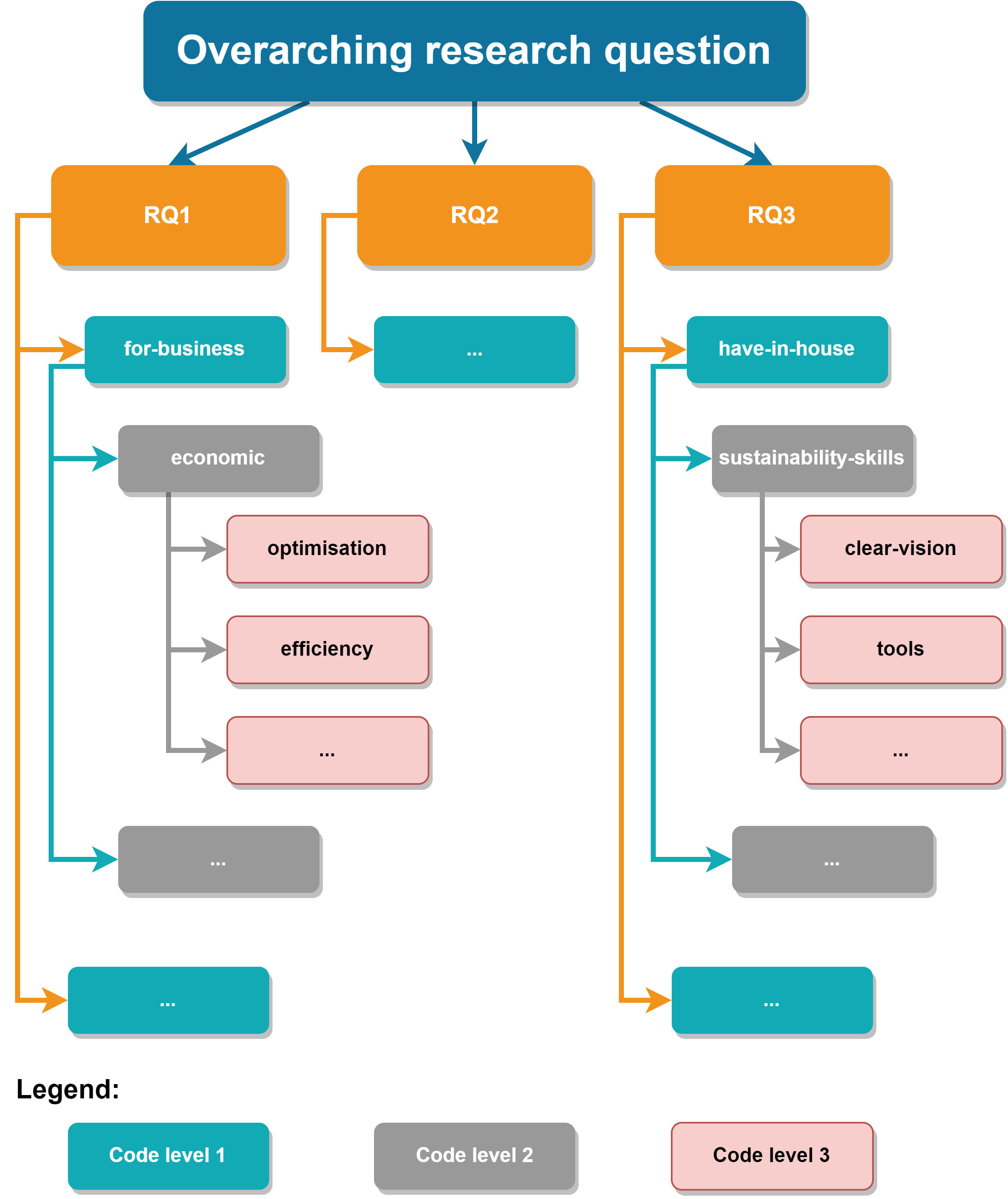}
  \caption{Structure of the codebook}
  \label{fig:codebook}
\end{figure}

The codebook has two purposes. Firstly, it formalises what the researchers have analysed from the data during the \textit{data extraction pilot} stage. Secondly, it is used as a guideline in the \textit{data extraction} stage, guiding the researchers who validate and correct the results initiated in the previous step. In the codebook, at the highest level, the coded data were organised according to our three research questions. The main topics are interests, target achievements, and competencies. The codes belonging to these main topics were further organised into three levels depending on their abstraction. The deeper level a code goes, the more detail it represents. 
The codebook\footnote{Codebook access link: \url{https://bit.ly/3wgt0dp}} is shared as supplemental material to help readers better understand the findings that we report in Section~\ref{sec:results} 
where the codes are highlighted in \code{bold}. 
The codebook is organised as follows. Horizontally, it is divided in accordance with the main topics mentioned above (i.e., interests, target achievements, and competencies). The codes belonging to each main topic are vertically distributed into three levels. Each code is accompanied by a definition, a description of when the code is applicable, and a coded text example with the respective source.
Figure~\ref{fig:codebook} illustrates the overall structure of the codebook and Table~\ref{tab:data_coding} shows sample final results of our data analysis phase, which can also be found in the supplied codebook. The examples contain quotes taken from the interviews and how they are coded during data analysis. The first set of two examples is for RQ1, which helps us identify why sustainability becomes an interest for our interviewed organisations in terms of economical aspects. The second set shows the organisations' goals in relation to sustainability (RQ2). The last set for RQ3 indicates skills that employees of our interviewed organisations possess in order to help them achieve the established sustainability goals.

\begin{table*}[h]

  \caption{Example results of data analysis}
  \label{tab:data_coding}
  \begin{tabular}{p{1.5cm}p{9.2cm}p{1.5cm}p{4.2cm}}
    \toprule
    \textbf{Interviewed} & \textbf{Quote} & \textbf{RQ} & \textbf{Code}\\
    \textbf{organisation} & &  \textbf{} & \textbf{} \\
    \midrule
        3 & ``\textit{Young people want to join us because they want to work for technologies which are basically improving sustainability.}'' & RQ1 & for-business / economical / attract-talents \\
        27 & ``\textit{If we have skills to optimise our algorithms to reduce our cloud costs, it is a financial benefit for us.} & RQ1 & for-business / economical / economic-return \\
    \hline
        1 & ``\textit{When you see the flow of value and start identifying where waste goes, you can deliver a better and cleaner system.}'' & RQ2 & goals / for-own-process / understand-to-help-customers \\
        18 & ``\textit{We are seeing IT more as a means to get insights on how to create sustainability, to be able to steer the effects and decisions you make.}'' & RQ2 & goals / for-own-process / add-value \\
    \hline
        2 & ``\textit{We have talents with long backgrounds in software development and good knowledge of building top-quality software. }'' & RQ3 & have-in-house / technical / quality software \\
        15 & ``\textit{We have a lot of employees who are educated in IT. They are also business advisors.}' & RQ3 & have-in-house / technical / business \\
     \bottomrule
\end{tabular}
\end{table*}


\section{Results}\label{sec:results}
This section describes the demographics of participants and the findings with regard to our research questions.

\subsection{Demographics: role with respect to sustainability} \label{sec:results:demographics:model}
In this section, we report the demographics concerning the business visions of our interviewed organisations and their perceptions of sustainability.

\subsubsection{Organisations' role with respect to sustainability}
We classified our interviewed organisations as producers or consumers (or both) of sustainability solutions in IT. 
The \textit{producers} are the organisations who produce tools or software solutions to support sustainability initiatives. The organisations that use these products are classified as \textit{consumers}. Some organisations may play both roles. The ``Business model with respect to sustainability'' column of Table~\ref{tab:interview_companies} shows the classifications of the organisations who use one of these models or both. While the majority of our interviewed organisations, 16 out of 28 organisations (55.6\%), from nine different countries are solely producers, only three organisations (11.1\%) from three countries are solely consumers. Nine organisations (33.3\%) from three countries play both roles. 
Finally, 9 out of 28 organisations operate in the public sector, while the rest are private organisations.

In our sample, many organisations develop sustainability solutions in-house rather than rely on other organisations. In many cases, the sector to which a company belongs does not impact the role adopted. Furthermore, the producer role is more common in software and technology organisations. Therefore, we can see that digitalisation plays a key role in providing consumers with sustainable solutions. Finally, we observe that organisations adopting both roles belong to the public sector (e.g., energy, water management) acting as end-users that demand sustainable solutions but also develop sustainability solutions in their IT department that are used by the environmental department.

\subsubsection{Organisations' perception of sustainability}\label{sec:results:demographics:perception}
We did not explicitly ask for the organisations' perception of sustainability during the individual interviews and focus group interviews as we did not want to have a confirmation bias, especially in the focus groups. 
However, we did extract the organisations' perceptions of sustainability that 
emerged when analysing the qualitative data. Overall, the main focus of organisations when discussing sustainability is on the environmental dimension. For eleven organisations, their statements can be interpreted that they perceived sustainability as environmental issues such as carbon emissions, climate change, and energy consumption. 

Eight organisations did also mention the economic dimension as part of sustainability. From these, three organisations referred to the financial impact of their products on their businesses (``\textit{[...] whenever we are making the systems, especially architectural or technological decisions, we consider the economical impact a lot}'' -- Org. 2) and also talked about economic sustainability in the sense of circular economy. The social and technical dimensions have been considered as part of sustainability by seven organisations. For the social dimension, the focus is on their own workforce (e.g., providing yoga classes), the customer (e.g., improving customer satisfaction) and society (e.g., fighting against poverty) as a whole.
For the technical dimension, the interviewees' statements suggest that sustainability is related to a quality attribute of IT products and services, such as reusability, robustness, security, etc. Finally, 12 organisations explicitly consider sustainability as related to more than one dimension and surprisingly, only four organisations mentioned the SDGs as relevant to them. In conclusion, we observe that a prominent focus of organisations when discussing sustainability is on the environment. Economic, technical, and social issues are also popular topics.

\subsection{RQ1: \textit{What are the interests of organisations with an IT division with respect to sustainability?}}\label{sec:results:rq1}
For this RQ we asked the organisations about their interests in sustainability from four perspectives: their business, their customers, their shareholders, and their stakeholders.


\subsubsection{Interests with respect to business}
When discussing the reason why sustainability is interesting for their \code{business}, it was not surprising to observe that economic reasons play an important role and are followed by moral and social matters.

With regard to \code{economic} reasons, sustainability helped our interviewed organisations open new business opportunities, increase their competitiveness, give them the license to operate, reduce costs, and acquire and retain talent more easily. In particular, sixteen organisations affirmed an interest in sustainability because it creates new \code{business opportunities} or helps mitigate potential threats. Overall, our interviewed organisations have one of the following three profiles: sustainability is their main business (e.g. they offer solutions for circular economy or sustainability reporting), they are in an industry that is being highly impacted by sustainability demands (e.g. mobility), or their customers are demanding it. Nine organisations viewed sustainability as a matter of \code{competitiveness and survival}. While some use sustainability to differentiate themselves from competitors, others are aware that other organisations are investing in sustainability-related initiatives and do not want to be left behind. For example, Org. 6 stated that ``\textit{[sustainability] is another point of differentiation}'' and Org. 23 mentioned that ``\textit{all the organisations that are emerging and that are effectively working, are those that have that sustainable consistency at an environmental and social level}''. Finally, for some organisations, it is a matter of making sure that they will continue to \code{utilise the resources} they need to function; for example, Org. 22 is fighting climate change because ``\textit{we are going to stop having the main resource of our factory, which is the water}''. Three of the organisations explicitly stated that implementing sustainable practices can bring them \code{economic advances}. Org. 27, for example, has a predictive algorithm for product demand that helps its clients minimize food waste and seeks to optimise its algorithms to reduce cloud and energy costs. For six of the organisations, sustainable practices were adopted to comply (or help others to comply) with \code{regulations}. Org. 22, for example, shared that \textit{``everything to do with climate- and  environmental policies have become structural''}. Finally, three organisations saw that sustainability is vital to \code{attracting talent}. These organisations feel that highly skilled professionals want to work for firms where they can share their values and put time and effort into something meaningful.

Sustainability was also related to \code{moral} concerns. Eight organisations invested in sustainability because they truly believe in it, sometimes being directly related to \code{aligning to the company's values}. Two illustrations of this belief are \textit{``our goal as a company has been focused on providing something to society and not just doing profit''} (Org. 18) and \textit{``our mission or reason for existence is that our business is coming from sustainability''} (Org. 3).

Unsurprisingly, when talking about their interest in sustainability, \code{environmental} concerns such as reduction of waste, water and energy use, and carbon emissions were the most present (mentioned by fifteen organisations). These concerns were related to both the purpose and the operation of the business.
Finally, four organisations explicitly 
stated that sustainability (or specific dimensions of it) was \code{not of concern} to them. 
For example, 
ecological sustainability was \textit{``almost the last perspective for us in day-to-day life''} (Org. 2).

\subsubsection{Interests with respect to customers} 
Differently from the business perspective (mainly focused on the economics), the \code{customers} of our interviewed organisations are reported to be most attracted to sustainability by moral values. In particular, they align their businesses to sustainability due to ecological and societal concerns. Economics, e.g., business opportunities returns, is the second most popular reason for investments. Here it is worth highlighting that several of the interviewed organisations adopted a business-to-business model, therefore, their customers were other organisations rather than individuals.

Among the \code{moral} reasons driving our interviewed organisations' customers to sustainability, sixteen organisations shared that their customers wanted to \code{protect the environment} by reducing carbon emissions and electricity use. At the same time, \code{social matters} were of concern to nine organisations. Especially, COVID-19 was the most frequently mentioned issue as the pandemic forced organisations to adapt their businesses for survival. For example, Org. 23 shared that most products requested by its customers in the last two years were related to the pandemic. In addition, four organisations viewed \code{value-alignment} as another reason for customers' interest in sustainability because this is an important concept in society. 

Regarding \code{economic} aspects, \code{investment returns} and \code{business opportunities} are the two most popular reasons. Three organisations mentioned that sustainability is a core business value of their customers: 
``\textit{Sustainability and circular economy are [our customer]'s core business.}'' (Org. 3). At the same time, the focus on sustainability has \code{evolved}, so our interviewed organisations and their customers had to proactively adapt their businesses to the new trends. 

Interestingly, seven organisations mentioned that despite having an interest in sustainability and in products addressing sustainability-related aspects, they still struggle to win customers due to \code{no interest}. Org. 21, for example, admitted that ``\textit{I'd love to design more services for sustainability, but I don't get any requests, and I really struggle to sell it.}''

\subsubsection{Interests with respect to shareholders}
As compared to the interest with respect to business and customers, the interest with respect to \code{shareholders} seems less important as it has been mentioned by only thirteen organisations. In particular, economic interests have been mentioned by nine organisations followed by societal concerns (four organisations). Three organisations mentioned that the interest of their shareholders in sustainability had changed over time. 
The \code{economic} interest is what organisations see as important for their shareholders. They mainly argue that their shareholders consider sustainability as a \code{business opportunity} to increase their financial performance and their market share. See, e.g., Org. 24: ``\textit{If we don't adapt the business and create new KPIs and processes related to sustainability, the risk is high for the shareholders because we can lose some parts of the market.}'' 

Four organisations do consider \code{societal} concerns as an important aspect for their shareholders. Org. 18 put it as follows: ``\textit{[shareholders] have a vision not just of making a profit but as a vision of contributing to a better society (...).}'' Additionally, the interviewees state that the shareholders' interest in sustainability \code{evolves} due to compliance constraints (e.g., EU taxonomy for sustainable activities\footnote{\url{https://bit.ly/3xYpBAF}}) and societal concerns (e.g., social responsibility).

\subsubsection{Interests with respect to stakeholders} 
The responses from the organisations show that sustainability interest from \code{stakeholders} is highly influenced by media news on sustainability, especially about \code{environmental concerns} like reducing emissions and fighting climate change: ``\textit{(...) if you take into account the ecological point of view, you'll have less $\mathrm{CO}_2$}'' (Org. 7). There are also several drivers for the \code{evolution} of stakeholders' interests in sustainability, such as the United Nations Framework Convention on Climate Change (UNFCCC), which launched the Race to Zero campaign and influenced several organisations working towards sustainability. Three organisations are working on building an ecosystem with partners who share the same sustainability values because sustainability \code{value-alignment} is an important element for their stakeholders.

In addition, \code{employees} are key stakeholders who drive sustainability interest within six organisations because they want to feel a sense of contributing positively towards sustainability: ``\textit{They are aware of the sustainability issues, and they want to have a positive impact on these issues.}'' (Org. 18). Also, organisations want to create employee satisfaction through different aspects of sustainability to attract talent based on the company's identity and activities towards sustainability. In addition for eight organisations, stakeholders' interests revolve around \code{societal concerns}, such as human rights, as well as individuals taking action and being accountable for their activities towards sustainability.  

\begin{mdframed}[style=mpdframe,everyline=true,frametitle=\color{white}{RQ1 summary},backgroundcolor=gray!15]
\textbf{Interests with respect to business:}\newline
The economic benefit brought by sustainability was the main driving force for the organisations' interest. Concerns about environmental impacts were also very present.\newline  
--------------------------------------------------------------------------\newline  
\noindent\textbf{Interests with respect to customers:}\newline
Our interviewed organisations felt that their customers had environmental and social concerns that they had to respond to.\newline  
--------------------------------------------------------------------------\newline  
\noindent\textbf{Interests with respect to shareholders:}\newline
Economic benefits are what the organisations see as most important for their shareholders.\newline  
--------------------------------------------------------------------------\newline  
\noindent\textbf{Interests with respect to stakeholders:}\newline
External drivers such as the media and the development of international frameworks highly influence the interest of stakeholders in the sustainability of the organisations. Another main driver is employees' personal interest.
\end{mdframed}

\subsection{RQ2: \textit{What do organisations want to achieve with respect to sustainability, and how?}}
To answer this RQ, we asked the organisations about what they want to achieve with sustainability in their business, how their ICT products/services support achieving these goals, and what difficulties they faced, or expect to face, in achieving these goals.

\subsubsection{Established sustainability goals}
The sustainability \code{goals} of the interviewed organisations focus primarily on their processes, followed by their need to create or improve their products, and finally on the external factors impacting their goals. 
Interestingly, social goals (e.g., human rights, and inclusion) are not their top priority at the moment. 
RQ1 findings show that the organisations and their customers have shown a strong interest in social issues when referring to sustainability. Particularly, social matters are the second highest interest of our 
organisations and of their customers. However, these aspects are not taken into account when organisations define sustainability goals. It indicates that although social matters are good reasons to draw organisations' attention to sustainability, they still do not create viable opportunities for them to set related business goals.


In relation to the internal working \code{process}, fourteen organisations highlighted how \code{process improvement} had contributed to addressing their sustainability goals, including automation and optimization. For example, Org. 1 highlighted the importance of values and mindset: ``\textit{shift [our business partners]'s mind to the value mindset from the project mindset.}'' Five organisations stressed the importance of \code{collaboration and leading by example} to inspire, influence, and motivate others to follow their lead. Six organisations reflected on their personal and professional decisions motivated by internal organizational and personal values to identify opportunities and take responsibility in relation to what motivates and positively influences their \code{decision making process}. Org. 18 stressed that as a company, they \textit{''have always had the goal of having a positive contribution to society as a whole''}. Four organisations commented that the organisational environment, belief, awareness, and communication linked to values were critical to \code{changing culture}. However, this needs to \textit{''be accompanied by a sustainable pace''} (Org. 1). Three organisations had introduced \code{new process}es to address sustainability internally but wonder how they could optimise their processes. 

Concerning the organisations' planned \code{products}, seven organisations highlighted opportunities to develop \code{new products} to create new markets. For example, the core business of Org. 15 is reviewing their clients' products in order to suggest new sustainable business strategies: ``\textit{We try to take a position in the market not as a regular IT supplier but more as a partner. 
We [and our customers] aim not only to make financial benefits but also benefit the environment and the social side.}'' Five organisations highlighted how they are improving product quality for both their clients and their organization by demonstrating how sustainability can be integrated as a core element. When clients lacked the required knowledge or expertise, these organisations were able to provide models and examples of good practice. 

Finally, regarding the \code{external} factors affecting the company goals, three organisations discussed the importance of their customers making the \code{right decision} for the larger community and global sustainability. Org. 2 even took this as far as challenging the need for a customer's development proposal, resulting in business loss when the potential customer decided to cancel the project. Also, three organisations highlighted the importance of looking beyond the boundary of the company to engage with sustainability in their \code{supply chains}. Org. 23 emphasised that ``\textit{we are going to review our supplier code of ethics. We want our suppliers to sign it and take responsibility because if we do it well, we have to go hand in hand with people who do it well.}'' However, they ponder how to achieve this, and need tools to help 
them in making the right decisions.

\subsubsection{Plan and experience for achieving the goals}
This section presents findings on executing sustainability goals and reported experiences.

\setlength{\parindent}{0pt}
\subsubsubsection{Plan}
Among the \code{steps} to achieve the established sustainability goals, organisations mention actions related to external factors with an impact on their goals, as well as the changes required to their internal process and product. 

\setlength{\parindent}{14,30342pt}
With regard to \code{external} factors, the most cited concern was seeking \code{collaboration}. Seven organisations prized collaborations with external entities (e.g., clients, municipalities, NGOs, and universities) and international alliances to push their limits, making them more ambitious in their goals. The UN Global Compact\footnote{\url{https://unglobalcompact.org}} was acknowledged as a good way to create synergies, gain strength, and be inspired by the work of others. The interviewed organisations also held thorough discussions on internal \code{process} transformation, focusing first on process design, second on tools, certification, and measurement, and third on implementation, evaluation, and internal collaboration. 

The \code{design} of sustainability processes was cited by five organisations, that use agile and incremental improvements, the ABCD process\footnote{ABCD is part of the Framework for Strategic Sustainable Development (SSD). Source: \url{https://www.naturalstep.ca/abcd}}, and the SSD framework. Org. 1, advocating for agile approaches, highlighted the importance of seeing a system as a matrix connecting its different parts to be aware of the effects of a given action on the system's value chain.
Four organisations highlighted the need for \code{tools} for sustainable systems. The tools mentioned are Care to Create, a flourishing business canvas\footnote{https://flourishingbusiness.org} to capture economic and social value, SSD framework for strategic sustainable development, and software for environmental accounting. Four other organisations discussed \code{certification}, sharing their uncertainties and concerns on measurement processes to achieve sustainability, particularly related to the lack of consistent methodologies to implement it. While two organisations referred to BREEAM and B Corp, one mentioned the importance of all types of certifications they adopted to prevent anti-corruption and comply with data protection. Another company reported having their Environmental programme approved and a Social Responsibility programme already draughted. 
Also, crucial to four organisations are the \code{measurement} processes to achieve sustainability, and related to this is the lack of consistent methodologies to implement such measurements, as stated by (Org. 10)
 \textit{''there's not a formal training program on how to develop sustainable solutions (...). [there's no] consistent methodology that guidelines how to implement it. (...) it will come to be certain because we're in a highly regulated company.''}
Org. 17 uses 
BREEAM and other certifications to measure and track the achievement of their goals. 

Three organisations mentioned the importance of process \code{evaluation}, either by setting clear objectives (e.g., becoming $\mathrm{CO}_2$ neutral), implementing and testing them or by defining end-to-end sustainable propositions to make contributions to the community. Org. 1 also complained that when a company is in financial trouble, the quickest decision is firing people instead of analysing the system and thinking of a way to add value to it. They called for a change of mentality where employees do not simply follow instructions but are able to raise their concerns.


Lastly, three other organisations proposed different ways to work in \code{collaboration} with their clients and partners to promote sustainability, from organising workshops to joining Global Compact and offering their clients solutions with extra sustainability features.
\newline
\setlength{\parindent}{0pt}
\subsubsubsection{Experience}
While achieving sustainability goals, the organisations collaborated with external business partners and reformed the internal organisation. Among the adopted actions, reducing energy consumption and cutting carbon emissions are the two most mentioned.

\setlength{\parindent}{14,30342pt}

Regarding \code{external} factors, three organisations reported that they sought \code{collaboration} with business partners to bridge the sustainability gaps within their organisations. Asking for training to enrich the workforce's knowledge about sustainability and raise clients' awareness of sustainability is one of the most popular choices: ``\textit{We have organised some workshops with an environmental expert and also invited clients to speak precisely about the importance of sustainability.}'' (Org. 23). Another solution is to purchase the services the organisations need to achieve their sustainability goals. For example, Org. 11 shared that they were working with a software consultancy company to produce a reporting toolbox tracking the amount of carbon emitted by its systems.

Regarding \code{internal} reform, organisations have adopted different solutions, such as \code{reducing carbon footprint} (seven organisations) and employing software technology to \code{automate} the working process (three organisations). To reduce carbon emissions, Org. 23 was particularly proud of its involvement in reforestation efforts to offset its carbon footprint. Org. 4 has recently provided bicycles for staff and encouraged cycling to work; the initiative has received high appreciation from its employees and media. Regarding automation, Org. 28 states: ``\textit{Test automation helps build sustainable software because you have confidence that a particular unit operates in a certain way.}'' Five organisations have re-\code{designed} their internal working processes in several ways to align themselves with the established sustainability goals. See the recommended procedure of Org. 1: ``\textit{So the first part is to identify value, to take care of value. Once you see that, you start changing things with little experiments.}'' Three organisations also reported their internal \code{culture} coincidentally changed on the journey to accomplish the goals. Org. 25 stated that aiming for sustainability goals initially created more work and caused objections, but over time, the passion has been growing among the staff in several departments.

Finally, for those who sell \code{products} related to sustainability, such as power-saving utilities and carbon emission reporting software, three organisations mentioned that they invested in new infrastructure to \code{reduce energy consumption}. In particular, Org. 24 stated: ``\textit{We sell different products and services that help reduce emissions.}''

\subsubsection{Encountered difficulties}\label{sec:results:rq2:difficulties}
The difficulties in achieving the established sustainability goals are categorised into external and internal. More internal difficulties were reported.

\setlength{\parindent}{0pt}
\subsubsubsection{External difficulties}

Economic barriers were the most frequently mentioned, followed by policy issues.

\setlength{\parindent}{14,30342pt}

Regarding \code{economic} barriers, four organisations emphasized the difficulty in finding \code{customers} willing to pay for more sustainable products or services. Org. 19 specifically mentioned that customers' procurement teams were ``\textit{only focused on price}'' and they had to sell the ideas to other decision-makers outside of procurement. Org. 16 aimed to be an enabler for the \code{circular economy} but was hindered by a lack of consistent models from other organisations. The economic barriers are not just in relation to customers; another company reflected on the difficulty of securing \code{investors}, who are focused on rapid growth and scaling up.

\code{Policy} issues were highlighted for not creating the conditions for sustainable products and services to compete. Policymakers represent a key difficulty, according to five organisations. If the policy context does not require or incentivise sustainability improvements in products and services, organisations struggle to compete against less sustainable alternative suppliers. Several organisations identified a gap between the aspirations politicians state and the \code{regulations} in force. Org. 9 felt this might be due to politicians' lack of understanding of digitalisation, who then ``\textit{proposed unreasonable laws}''. The impact of these policy failures was reflected in the second most serious concern, that customers were not prepared to pay more for these sustainable products and services. Four organisations cited external \code{technological} barriers, such as a lack of charging infrastructure for electric vehicles. 

\setlength{\parindent}{0pt}
\subsubsubsection{Internal difficulties} People and processes are overwhelmingly represented in the organisations' answers, while technologies are barely mentioned. Specifically, they appear 25, 23, and 2 times, respectively. 
It shows that non-technological difficulties are prevalent in the industry. 

\setlength{\parindent}{14,30342pt}
 
Regarding \code{people} barriers, 14 organisations pointed out the lack of \code{understanding} of sustainability concepts as one of their most significant challenges. This may be due to the extent and vagueness of the concepts themselves and the insufficient knowledge of their employees. Some further explained that the complex conceptualisation of sustainability at a company level also makes searching for sustainability skills in new potential employees a challenging task, as the existing workforce is not qualified to assess the needed skills or their fulfillment by applicants. Org. 4, for example, stated: ``\textit{Terminology/concept is still vague, especially what kind of skills and competencies you already have in-house.}'' Furthermore, ten other organisations identified the \code{culture} of their employees, its complexity and inertia, as one of the important internal challenges. Org. 25 states that ``\textit{We had many key stakeholders and people at the bottom. We had some commitment on the high management, but it didn't connect because it was blocked in sub-optimisations of middle management who try optimising their own budget or their own question.}'' The culture is often oriented toward different KPIs (Key Performance Indicators) and conflicts with sustainability goals. The short-term priorities and missing skills of \code{decision-makers} have also been mentioned by seven organisations. Org. 18 noted that ``\textit{managers usually have more short-term goals and it's not easy to sell them a long-term plan that won't give profits to the company for years.}'' 

With regard to difficulties arising from the internal working \code{process}, we find a \code{financial} trade-off between short-term financial profitability and long-term sustainability goals one of the most frequently mentioned difficulties (encountered by 15 organisations). The issue occurs in both our interviewed organisations and their customers. Nine organisations encountered an issue regarding the ability to carry out sustainability-relevant \code{measurements}. For example, Org. 9 admitted that ``\textit{[their employees] don't know how to measure sustainability, or how to advance the policy agenda on sustainability using IT or digitalisation}''. Org. 4 highlighted the challenges to calculate the $\mathrm{CO}_2$ footprint of cloud services.

At first sight, these external economic barriers and internal short-term financial gains may seem to contradict the results of RQ1, which states that the economic benefits brought by sustainability are the main driving force for the organizations' interest in it. This is a common conflict. Sustainability demands long-term investment. However, it can be difficult to convince internal and external stakeholders to sacrifice fast economic growth in favour of the long-term economic benefits brought by sustainability.
\begin{mdframed}[style=mpdframe,everyline=true,frametitle=\color{white}{RQ2 summary},backgroundcolor=gray!15]
\textbf{Established sustainability goals:}\newline
The interviewed organisations highlighted the need for improving their design processes and  products to support sustainability and stressed the importance of a change in culture to positively contribute to society and help their customers make the right decisions. \newline  
--------------------------------------------------------------------------\newline  
\noindent\textbf{Execution plan and experience to achieve the goals:}\newline
To achieve the established sustainability goals, the organisations focus on seeking collaboration with their business partners and other external entities, transforming their internal working processes, and developing tools to support interconnectivity, interdependence, and adaptability. The experiences reported include knowing how to collaborate with external stakeholders effectively, reducing carbon emissions, and applying automation when possible.\newline  
--------------------------------------------------------------------------\newline  
\noindent\textbf{Encountered difficulties:}\newline
The difficulties reported are caused by internal and external factors. The major internal factors are those related to the trade-off between short-term financial profitability and long-term sustainability goals, an unclear understanding of sustainability concepts and goals, and the culture of the employees, which is often oriented towards KPIs in conflict with sustainability goals.
With regard to external factors, economic barriers and inadequate policies are the two most frequently mentioned.
\end{mdframed}

\subsection{RQ3: ``\textit{What are the sustainability-related competencies and skills that organisations need to achieve the established sustainability goals?}''}\label{sec:results:rq3}
To answer this RQ, we asked the respondents about the skills and competencies available and missing within their organisations, as well as their approaches to acquiring those identified as lacking.
In our context, skills are the specific learned abilities needed to perform a given job well. On the other hand, competencies are the person's knowledge and behaviours that lead them to be successful in a job.

\subsubsection{Skills and competencies available in-house}\label{sec:results:rq3:having}
From the interview data, we observe that there is a wide variety of skills and competencies required by our interviewed organisations to achieve their established sustainability goals. Only a subset of the skills and competencies were claimed to be available in our respondents' organisations but not all. In addition, the sets of available skills and competencies are not the same among the organisations. 

These skills and competencies have been categorised into sustainability-related skills (e.g., organisations state that they have knowledge of the sustainability-related regulations in their domain), soft skills (e.g., organisations see that they have good problem-solving skills in sustainability challenges), and technical skills (e.g., organisations see they have the ability to create technically sustainable solutions). 

Organisations had two different perspectives on \code{sustainability-related} skills. They either thought about sustainability in a holistic, higher-level manner or focused on some specific details of sustainability. From a high-level perspective, many of our interviewed organisations believed that sustainability knowledge is not that important for IT staff. In particular, seven organisations thought that IT staff do \code{not need} to acquire sustainability knowledge, and four other organisations required \code{little background} from their employees. These organisations do not mean that they do not emphasise sustainability but they distinguish IT people from sustainability experts. When the organisations focused on specific sustainability-related skills, they mentioned different sustainability dimensions, application domains, and tools and approaches to achieve them. The organisations seem to have an understanding of both \code{social} and \code{environmental} dimensions: ``\textit{[Our staff] have the environmental knowledge and have involved in GRI\footnote{GRI: an international independent standards organisation that helps businesses, governments and other organisations understand and communicate their impacts on issues such as climate change, human rights, and corruption. Source: \url{https://www.globalreporting.org/}} and CDP\footnote{CDP: an international non-profit organisation that helps organisations and cities disclose their environmental impact. Source: \url{https://www.cdp.net/en}}. They can use these competencies to help our customers}'' (Org. 14). These dimensions also link closely to the \code{regulations} that the organisations have to obey, as mentioned by Org. 14: ``\textit{we've been working with our customers and see how EU regulations have evolved}''.

The organisations pointed out several \code{soft skills} they think are valuable while aiming for sustainable solutions. Some of these skills are rather traditional, like \code{problem-solving} and \code{collaboration}, while others, such as \code{common-sense}, \code{reflection and influencing} relate to the aim for effects on sustainability. The problem-solving and collaboration skills link closely to the sustainability-related skills presented above. 
The most referenced (seven organisations) category of soft skills was \code{influencing}. It shows how organisations recognise their skills to influence the customer and the outcomes. For example, Org. 2 mentioned, ``\textit{...their daily work has influenced customers' teams about what should be done}''.

The last set of skills that the organisations seek to have in-house is of a \code{technical} nature. Most of the categories link with IT-related skills and were clearly stated by our interviewed organisations, for example, \code{software quality}, \code{user-centricity accessibility}, \code{architecture}, \code{data management}, and \code{systems thinking} skills. While the first four skills are more familiar to the ICT community, the meaning of the systems thinking skill can be explained by the following statement of Org. 18: ``\textit{we are software engineers, so systems are quite familiar to us, so we know the different parts of it, how they interact, and how they join together.}'' Surprisingly, our interviewed organisations emphasised most (five references) the business skills they have in-house and considered part of the technical skillset. This finding is highlighted by Org. 15: ``\textit{Most of our developers are also business advisors. We are now adding more competent business advisors having an IT background.}''

\subsubsection{Skills and competencies missing in-house}\label{sec:results:skills:missing}
Similar to the analysis for skills and competencies available in-house, in this section, we cluster the missing skills and competencies into three categories: sustainability-related skills and competencies, soft skills, and technical skills. 

With regard to \code{sustainability-related skills and competencies}, our interviewed organisations mentioned that they lacked many of them, such as the right talent who have knowledge of sustainability and can transfer that to new IT business opportunities, the IT staff who are both excellent at technical skills and have sustainability knowledge, and the talented programmers who can deliver energy-efficient code. In particular, six organisations recognised the importance of sustainability knowledge but faced difficulties in hiring the \code{right talent} with suitable sustainability-related skills. Five organisations mentioned that their staff lacked a \code{multi-disciplinary} skill set. For example, Org. 18 expected their IT staff to have some sustainability knowledge and stated that ``\textit{...would be good if we can have some of those professionals that could combine sustainability and good background on ICT.}'' Three organisations wanted their IT systems to be \code{energy-efficient} and \code{environment-friendly} but do not have developers who have these relevant programming skills.

Regarding missing \code{soft skills} that are crucial for organisations with respect to sustainability, communication, and systemic thinking were frequently mentioned. In particular, poor \code{communication} skill is an issue experienced by four organisations. This problem is visible for the people who work directly with customers (e.g., the marketing department): ``\textit{We have been facing the challenges that at least we think that we have the perfect idea that the customer would benefit from, but we are having difficulties selling that to the customers.}'' (Org. 4). The company also experienced that communication was ineffective among its IT staff. \code{Systemic thinking} is an ability to have a holistic view of factors and interactions that could contribute to a better possible outcome. However, three organisations could not identify this competence in their IT workforce. Org. 17 stated that ``\textit{if we had a framework or skills on how to put it all together in the bigger picture, we could have optimised our solutions for the entire system, not just specific code segments, or applications.}''

Software engineering is a technological field, but our respondents mentioned several missing \code{technical skills}. Specifically, six organisations reported the lack of \code{metrics} to measure the impact of their IT products on sustainability. For instance, Org. 2 stated: ``\textit{We don't have good means to measure the sustainability level of certain software entities or our customers.}'' Data has been increasingly collected in recent years, but three organisations did not equip themselves well with \code{data management} skills. These organisations faced some difficulties in complying with GDPR (General Data Protection Regulation) in terms of data handling.

\subsubsection{Solutions to acquire what is missing}
Based on the identified skills and competencies missing in-house, we further investigated how the organisations are acquiring them. Overall, the acquisition strategies can be classified into two types: internal (i.e., carried out entirely within the company) or external (i.e., when the skills and competencies are provided from a source external to the company).

In relation to the \code{internal} approaches, the most common (mentioned by 20 organisations), unsurprisingly, is by providing and/or organising \code{in-house training}: ``\textit{...what we do to make the change in the organisation, I think we do both retraining and changing the behaviour in the organisation itself}'' (Org. 10). \code{Hiring} is also a widespread internal strategy, being mentioned by 13 organisations. Organisations use recruiting as an instrument to bring in new employees with suitable sustainability-related skills and competencies. This process also involves internal training in order to adapt newcomers to the organisation's culture and working process. There are several hiring targets being adopted by the interviewed organisations, including looking for specific pre-defined competencies (e.g., ``\textit{We hire people that have some specific set of skills and also have a passion or interest in sustainability}'' - Org. 20), people with the right mindset for the organisations. In addition, to address the communication issues, new hires are expected to ``\textit{...establish and maintain good discussions with customers and stakeholders}'' (Org. 2). Establishing \code{mentorship} programmes that engage experienced employees in sharing their own experience, knowledge, and know-how with other staff members; and conducting \code{internal events} for knowledge sharing are two other solutions mentioned by two organisations.

When it comes to \code{external} approaches, collaborating with universities, sending employees to participate in courses about sustainability, and hiring consultants are popular solutions. Firstly, we found that a significant number of organisations (11) expected to acquire the missing competencies either via new hires contributing the right background from their \code{university} education or by means of research collaborations with universities. For example, Org. 7 stated: ``\textit{To get this knowledge into our own company, we really need research or try to get information from universities.}'' Secondly, four other organisations frequently paid for \code{external courses} to train their own workforce. This strategy assumes that either suitable courses are available or external training organisations offer customisable course packages. Lastly, four organisations preferred to hire \code{consultants} on sustainability when sustainability-related competencies or skills are needed: ``\textit{For externally-reused software, cloud services, or to address specific sustainability goals, [...] we would partner up with somebody or buy consultancy hours}'' (Org. 16). This is another external strategy where missing competencies and skills are acquired temporarily and typically project-specific.
\begin{mdframed}[style=mpdframe,everyline=true,frametitle=\color{white}{RQ3 summary},backgroundcolor=gray!15]
\textbf{Skills and competencies available in-house:}\newline
In order to reduce the expectation level for the staff, many organisations separate IT departments from sustainability experts, so a sustainability background is not or little required for IT-skilled employees. However, specific soft skills (e.g., problem-solving, collaboration) and technical competencies (e.g., architecture, data management) are expected and available within the IT workforce to achieve the target sustainability goals.\newline  
--------------------------------------------------------------------------\newline  
\noindent\textbf{Skills and competencies missing in-house:}\newline
Despite separating IT departments from sustainability experts to reduce the need for sustainability knowledge, many organisations still want to fill that gap for their IT staff. Improving communication efficiency and defining sustainability measurement metrics have been often mentioned as missing soft and technical skills within the organisations' workforces.\newline  
--------------------------------------------------------------------------\newline  
\noindent\textbf{Solutions to acquire the missing skills and competencies:}\newline
The organisations have taken both internal and external approaches to fill sustainability knowledge gaps for their IT staff. Popular solutions are organising in-house training courses, collaborating with universities, sending employees to externally organised courses, and hiring sustainability consultants.
\end{mdframed}

\section{Interpretation of results}\label{sec:interpretation}

This section discusses the skills gaps apparent from our results 
and that future education programs should address. Then, Section \ref{sec:discussion} proposes concrete topics to be considered in those educational programmes and on-job training and collaborative activities in industry.

 
 
\textbf{(On RQ1) Interests in sustainability are diverse and evolving. 
Currently, professionals are not able to understand and relate multiple aspects of sustainability and translate these relations into concrete business when needed. Educational programmes should enable this competency while being flexible and ready for changes.}
 
\begin{figure}[h]
  \centering
  \includegraphics[width=\linewidth]{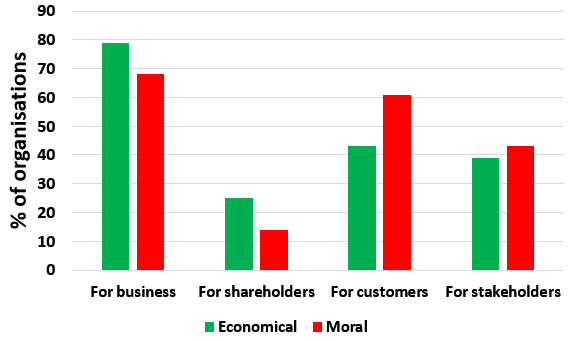}
  \caption{Reasons for the interest in sustainability}
  \label{fig:interests_sustainability}
\end{figure}

Our data show that economics is the main driver for our interviewed organisations and their shareholders to invest in sustainability. This is not surprising since they must survive in the market. However, as shown in Figure~\ref{fig:interests_sustainability}, there is pressure from customers and stakeholders to push for social and environmental sustainability impacts. 
So, one must try to turn these aspects
into economic profit; 
therefore, ideally, one must change the value proposition.
The business vision of an organisation must be decided based on multiple factors, its own business', customers', shareholders', and stakeholders' interests. As such, to prepare a better workforce for IT organisations, future educational programmes should help professionals relate concerns belonging to different sustainability dimensions and be able to translate such relations into concrete business plans. 

Around 40\% of the organisations mentioned that their interests in sustainability evolved due to new demands from customers and regulations. Such evolution might pose difficulties to organisations, so future educational programmes should be flexible and ready to adapt to changes.
At the same time, 14\% of the organisations are not yet concerned with sustainability. 
In this case, education plays a role in creating awareness within businesses about the relevance and opportunities of sustainability within IT, accelerating the IT industry's interest in sustainability. 

From the data, we identify that around 10\% of IT organisations are interested in sustainability but do not know how to take it into account. Educational programmes and specific industry training can help. 

\textbf{(On RQ2) 
%
The IT workforce needs sustainability-related competencies and a deeper understanding of how sustainability impacts the development processes and the resulting products. Thus, educational programmes should provide them with the tools to develop sustainability-aware systems and inspire others across their network of collaborating organisations to embrace sustainability initiatives.
}


Sustainability is activating organisations to change internally (e.g., how development and decision-making processes are revisited to address sustainability goals) and externally (e.g., how organisations offer their customers new or improved products and services with respect to sustainability). At the same time, organisations encounter a number of difficulties (further analysed below) that designers of future educational programmes for IT students should take into account when making improvements.
 
\begin{figure}[h]
  \centering
  \includegraphics[width=\linewidth]{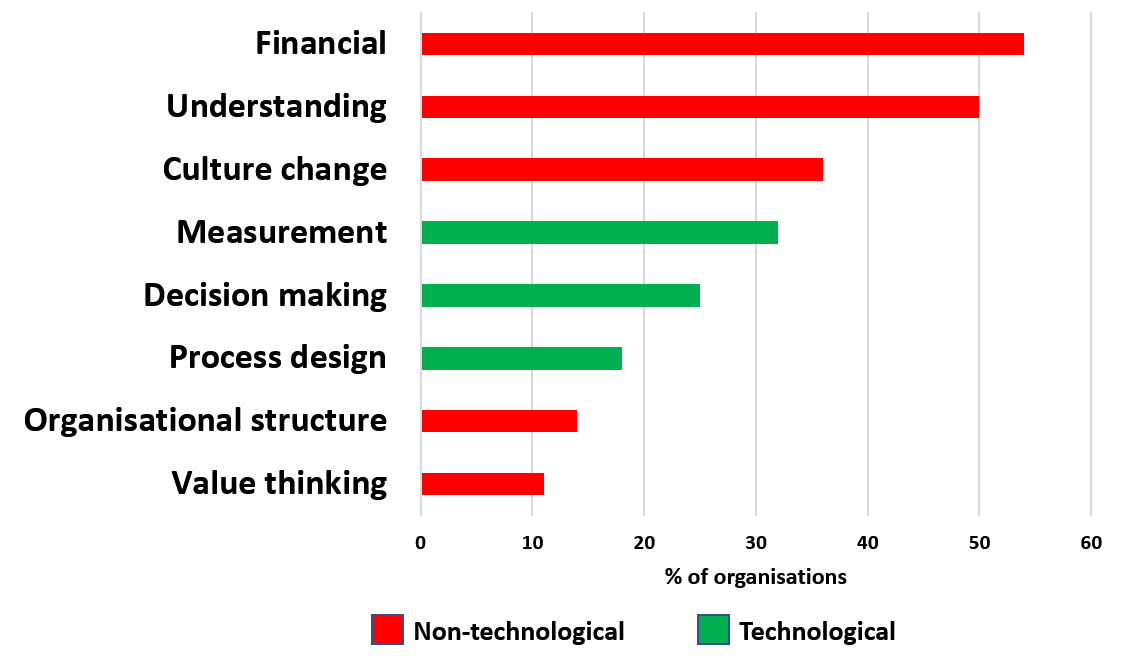}
  \caption{Internal difficulties encountered by organisations}
  \label{fig:internal_difficulty}
\end{figure}

As summarised in Figure~\ref{fig:internal_difficulty}, lack of funds and sustainability understanding, as well as the need to change the internal culture to favour sustainability more, were the three major internal difficulties reported by the organisations. Financial challenges may force organisations to downgrade sustainability objectives to survive, even though sustainability is something most of them want. 
This shows a need to create more value for sustainability-aware software products, and this may require 
a change of regulations to support sustainability-aware systems. 
Half of our interviewed organisations complained that their IT staff or colleagues need a better understanding of sustainability.  
%
At the same time, we observe that 20\% of the organisations consider current policies as insufficient for driving sustainable development, and 16\% struggle to persuade customers to pay more for sustainable products. This shows that even though customers want to buy more sustainable products, they are not necessarily willing to pay more. This is where politics can play a vital role in enforcing incentives, such as reducing taxes on green products and putting in place laws and adequate regulations. 
%
While non-technological issues seem to worry organisations more, they also drew our attention to the need for sustainability-related improved metrics, design processes and tools.


 \textbf{(On RQ3) 
 More sustainability skills and competencies are needed due to their rising importance in our daily life.
 Future educational programmes must be built upon three pillars: technology, soft skills, and sustainability knowledge.}

As educators, we believe that IT and sustainability are disruptive forces in today's society and will increasingly converge~\cite{business2017better}. 
Education programmes that give IT professionals strong soft- and sustainability skills will ease this process, both because it will set a basic common ground for collaboration among experts in both fields and because it will encourage every technical system to be built with essential sustainability characteristics in place. 

\begin{figure}[h]
  \centering
  \includegraphics[width=\linewidth]{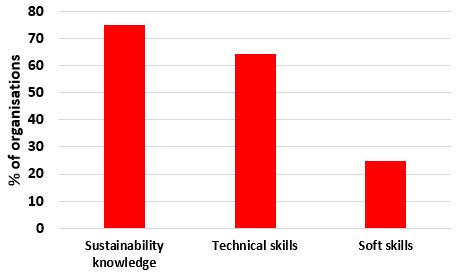}
  \caption{Skills and competencies missing in organisations' IT workforce}
  \label{fig:missing_skills}
\end{figure}

Many organisations have recognised the need for sustainability skills and competencies for their IT staff. As shown in Figure~\ref{fig:missing_skills}, the majority of our interviewed organisations agree that their IT workforce lacked sustainability knowledge and understanding (75\%), and technical skills to implement sustainability (65\%). Also, one-quarter of the interviewees pointed to the need for additional soft skills. 

Based on our observations, weaknesses regarding sustainability skills and competencies of the current IT workforce can lead to (1) difficulty in understanding sustainability in the business context, including sustainability strategies, approaches, and tools to support sustainable business models, (2) difficulty in translating business requirements into IT products and services with sustainability considerations, and (3) poor communication and soft skills, which is a classic problem with software engineers and programmers. 



\section{Discussion}\label{sec:discussion}



An increasing number of Education Institutions (HEI) recognise the need to integrate Sustainability  into their existing courses. The integration process is challenging~\cite{fiselier2018critical}, but there is guidance available, such as the UK QAA/HEA guidance on incorporating Education for Sustainable Development into the curricula. 
However, it is hard to identify guidance on developing skills for IT courses. 
To help, we developed a classification of the topics 
that should be considered when designing a curriculum for future courses in sustainability. This classification is the result of the interview findings. The classification is not meant to be complete, but in our opinion, it points to the most important topics based on our years of research and teaching in the area. In the following, we describe the importance of each topic, how it relates to our interview data and suggestions for how to consider it in future education.

\subsubsection*{Core sustainability knowledge}
Sustainability must be seen as a prerequisite for any IT product or service. We observed in our interviews that there was no clear consensus about what was meant by this term. If IT professionals do not adequately understand the main concepts of sustainability, they are likely to: keep looking at it as a trade-off or a nice-to-have; struggle to collaborate with sustainability experts, and might not fully see the value of it; shy away from public debate on technology and sustainability, not motivating policy changes; see it as a complementary skill to IT. Here, education could play a fundamental role in providing knowledge of the basic principles, concepts, and models, particularly, definitions and scoping of sustainability, clarification of persistent misconceptions and myths around sustainability facts, current statistics, fundamental concepts (e.g. dimensions of sustainability), as well as different models to explain sustainability in a certain context (the doughnut model~\cite{raworth2012safe}, the nine planetary boundaries~\cite{rockstrom2009planetary}, and orders of impact~\cite{hilty2015ict}.)

\subsubsection*{Systems thinking}
Systems thinking 
is a fundamental perspective change for being able to grasp sustainability issues. The focus on the overall big picture with its relationships between the stakeholders and systems, and their dynamics and balances --- instead of traditional divide and conquer approaches that are taught in engineering --- allows for the necessary shift in looking at a situation. Systems thinking is a powerful tool and mindset shift that enables a holistic view of sustainability across its different dimensions. Through the study, we observed that there were just a few organisations possessing this kind of competency within their workforces; for example, Org. 16 champions the value of recycling by producing food for fish from waste. In that case, the organisation goes beyond its main operation's boundary to seek ways to positively affect the natural environment. As educators, we are aware that systems thinking is not easy to teach and practice since it requires a good deal of domain knowledge. Yet, the potential consequences if this matter is not taken into consideration can be much greater, leading to cascading (potentially negative) impacts in different dimensions of sustainability. 
For example, if a country aims at building more and larger hyper-scale data centers addressing the demand for IT products and services, we must also consider the related societal implications for, among others, the (competing) demand for water for citizens, of renewable energy resources for cities, of land for agriculture - hence going beyond the specific ICT sector.


\subsubsection*{Soft skills}
Communication is an essential soft skill for the IT workforce since managers, peers, and clients have to exchange ideas on a daily basis. It is especially important to know how to communicate in a positive way, e.g. as in non-violent communication~\cite{rosenberg2002nonviolent} or compassionate communication that seeks to understand the motivation of the communication partner before attempting to get the point across. We believe soft skills should play a much more significant role than today in software engineering education programmes. This is confirmed by 12 of our 28 interviewees, who emphasised the importance of soft skills within their IT workforce.

Although the importance of soft skills is increasingly recognized in engineering degrees, educators often struggle to fit the topic properly into their classes. This difficulty may arise from several reasons. Sometimes, engineering teachers do not have the right knowledge and tools to effectively teach soft skills to students. On other occasions, they may feel that the time they have to cover the technical curricula is already too short, resulting in a weak integration of soft skills into their teaching (e.g. without giving the students the required time to reflect on and practice their soft skills).

We suggest designers of engineering programs should collaborate  with other courses where soft skills play a more critical role, such as management, leadership, and social psychology, to find more effective ways to integrate them into the curricula. A complementary activity could be to invite experts from industry to teach soft skills to IT students. This kind of collaboration will not only give students more practical perspectives but can also strengthen relationships between academia and industry and motivate students on the importance of such skills.  
\subsubsection*{Technical sustainability}
Technical sustainability refers to the capacity of the software system to endure in changing environments~\cite{becker2015sustainability}. Software systems are sustainable if they can be cost-efficiently maintained and evolved over their entire life-cycle~\cite{Martin2017}. Technical sustainability can be achieved through software architecture as it lays the foundation for the successful implementation, maintenance and evolution of sustainable software systems in a continually changing execution environment by providing a mechanism for reasoning about core software quality requirements that contribute to sustainability as a first-class, composite software quality. Addressing software sustainability at the architectural level, allows the inhibiting or enabling of systems quality attributes, reasoning about and managing change as the system evolves, predicting system qualities, as well as measuring architecturally significant requirements. However, the ability to determine sustainability as a core software quality of a software system from an architectural perspective remains an open research challenge, and existing architectural principles need to be adapted and novel architectural paradigms devised. In addition, there is a pressing need for new tooling to fit today's emergent and dynamic environments, where software systems are explicitly designed for continuous evolvability and adaptability without creating prohibitive architectural, technical debt. The skills needed to develop and estimate technical sustainability require training in software architecture, tools, and metrics to evaluate technical sustainability for the diversity of application domains. Whilst engineers may understand the importance of technical sustainability, metrics and dashboards of technical debt are less in evidence. For instance, Org. 18 mentions they use some metrics, but they lack automatic processes to evaluate the quality of systems. 
This is precisely one of the aspects in which technical sustainability can help to produce more sustainable systems.

\subsubsection*{Building the business case for sustainability}
Sustainability and the SDGs provide enormous business opportunities to organisations \cite{business2017better}. For example, Org. 2 stated quite frankly "\textit{Why we are so interested? [...] it's money.}"  In the short- to mid-term, companies can benefit from creating new systems to exploit these business opportunities, or from making their own systems 
more sustainable or, at least, more environmentally aware, appealing to increasingly sustainability-conscious customers and business partners. For example, the Environmental, Social, and Governance (ESG) framework is now used by stakeholders to evaluate how organizations manage risks and opportunities that relate to sustainability issues. Therefore, IT professionals need to understand better what drives the businesses they work for, the opportunities that focusing on sustainability open to businesses in general, and the threats faced by businesses causing harm to the environment and society. Understanding this might help them to champion the idea of sustainability internally and justify it in terms of economic, environmental, and societal reasons. Furthermore, in the mid- and long-term, companies should aim to create a positive, or at least non-negative impact, on all sustainability dimensions, independently of the purpose of their systems. 
This evolution of companies provides many opportunities for educators. The traditional practice-based courses such as capstones and hackathons could use these real-world challenges companies have and, as an outcome, provide possible solutions for companies. The challenges could be approached on different levels, from setting the company values and objectives to the design of the actual internal (software) development processes. In general, educators may collaborate with companies to increase practitioners' awareness of the possible sustainability impacts of their products and activities.

\subsubsection*{Sustainability Impacts and Measurements}
How companies improve their businesses by adopting sustainable practices requires specific Key Performance Indicators (KPIs) and metrics that can measure SDG targets or GRI indicators. While some domains (e.g. energy, transportation) count with standardized metrics to evaluate the sustainability of a solution, others do not. Consequently, many organizations have difficulties estimating sustainability and they need to define their metrics and sustainability indicators. To do so, IT staff and managers need to be trained on current metrics/KPIs and on how to create their own, such that organizations can be clear about their sustainability achievements. There are approaches and tools for assessing, quantitatively and qualitatively, sustainability impacts. They include techniques like scenario mapping, future backcasting, the SDG impact assessment tool~\cite{chalmers2019sdg}, sustainability assessment tools like the SAF Toolkit \cite{SAFToolkit,Lago2019}, and sustainability awareness frameworks like SusAF~\cite{duboc2020requirements}. Universities are being increasingly ranked according to sustainability (e.g. Times Higher Education Impact  Rankings), so as organisations adopt different sustainability goals they need different kinds of sustainability metrics\footnote{\url{https://www.timeshighereducation.com/impactrankings}}.

From our analysis of companies, we found evidence of this lack. For instance, Org. 2 highlights the lack of such metrics to understand the direct impacts of the product/system adopting sustainable solutions. In addition, Org. 27 observes that there is a lack of awareness on whether the solutions adopted are sustainable enough, partly because they do not have metrics. These and other technical challenges aimed to calculate the carbon or energy footprint (e.g. Org. 4) of sustainable solutions are why they demand specific training on well-defined KPIs to estimate the impacts of different sustainability initiatives to justify the efforts and expenses.

\subsubsection*{Values and Ethics}
Ultimately, values and ethics are fundamental concerns to making the world a fair and equitable place \cite{shearman1990meaning}. While ethics are culturally agreed-upon moral principles, ``values make no moral judgment''\cite[p.~113]{8693084}. Currently, our society relies on software systems to communicate worldwide and operate utilities providing the basics for human life (e.g. complex medical machines, nuclear plants, and electrical grids). Such systems focus on functionality and quality attributes, such as usability, availability, and security, but they do not respect our core values, such as social justice, transparency, or diversity \cite{schwartz2007basic}. Sustainable systems, however, should be aligned with core human values.
In this context, it is important that IT professionals are guided by a clear code of ethics, such as the ACM code of ethics\footnote{https://ethics.acm.org/code-of-ethics/software-engineering-code/} to produce socially and environmentally responsible systems. We call for ethics to be a standard part of software engineering.

In what concerns values, user-centred design, user-experience design, and values-sensitive design tackle more than typical software qualities, but they are still far from addressing core human values~\cite{Alidoosti2022}.
Value-driven methods, known in HCI and information systems, can be used in business analysis and requirements engineering, but they offer little guidance for the later stages of development. Some emerging works take a human-values view (e.g., GenderMag \cite{burnett2016gendermag} used to discover gender bias in software; or Alidoosti et al.~\cite{Alidoosti2022b} incorporating ethical values in software design), but more is still required to address human values systematically.
The good news is that software development methods could be adapted to handle human values. For example, the Karlskrona Manifesto on Sustainability Design\footnote{https://www.sustainabilitydesign.org/karlskrona-manifesto/}, or participatory design techniques can be taught to ensure that end-user values are taken into account.
Over 57\% of the interviewed organizations reveal that their customers and stakeholders want to protect the environment and almost 30\% are interested in focusing on sustainability due to moral concerns and social matters, resulting in the need for sustainability-value alignment of their business.

\subsubsection*{Standards and Legal aspects} 
The topic of legal aspects includes standards that may be required to be taken into account or that allow for certification, like the ISO 14000 family on the environment or the ISO 26000 standard on corporate social responsibility, or the LEED standard for buildings. It also includes issues of compliance and compliance assessment, the development of sustainability cases (think safety cases but for sustainability), and the impacts of newer and upcoming laws from privacy and information safety to software liability and their impacts on sustainability. One of the reasons for the slow adoption of sustainability initiatives in business has been the lack of mandatory requirements for action or reporting. Some interviewees mentioned specific regulations such as Waste Electrical and Electronic Equipment (WEEE), General Data Protection Regulations (GDPR) and Employment law (Org. 9, 11, 20, 23). There is a need for educators to be clear about the difference between what is required by law/regulation in different jurisdictions such as the EU Green Taxonomy reporting or employment protection, and what is voluntary that businesses may choose to do such as using the Global Reporting Initiative framework for sustainability reporting. 

\subsubsection*{Advocacy and lobbying}  
This topic raises the question of how impartial and neutral researchers and educators should be versus how much involved in advocacy and lobbying. We are in favour of taking a stance while allowing discussion space for all perspectives on an issue. One should not wait for regulation to start acting on sustainability. Regulation is often a late follower of social trends and is highly influenced by them. The last decades have witnessed several social and business movements towards sustainability, such as Corporate Social Responsibility, Fairtrade, Slow Food, Greenpeace, Natural Capitalism, B-corporations, etc. 
Education can play an important role in promoting and shaping movements such as the above. Universities should train future IT professionals to combine their technical and sustainability expertise to become strong advocates for sustainability. Therefore, curricula should also include tools for effective and positive advocacy in organizations, media and legislation, as well as lobbying. The importance of offering expertise to policymakers has been highlighted by one of our interviewees (Org. 9), who said: ``\textit{a lot of policymakers don't have a clue on digitalization matters (sic.), and because of that, they don't know what they're doing while writing the law.}"

\section{Threats to validity}\label{sec:threats}
We discuss our reasoning concerning the validity and limitations of this study by following the scheme described in 
~\cite{runeson2009guidelines}.

\textbf{Construct validity.} This aspect is related to whether during the interviews we asked the right questions according to our investigation objectives. To mitigate this threat, we formulated the questions by leveraging the background knowledge of the involved researchers, which indeed have experience in these types of research in software engineering in general, for at least ten years, and in software engineering related to sustainability for at least five years.

\textbf{Internal validity.} This is related to how we executed the study internally and analysed the results to avoid unknown factors that may influence the study's outcome. To mitigate this threat, we have taken the following actions. First, to improve the instrument (i.e., interview guide) used in the study, we spent time discussing the interview questions to ensure they covered our stated research questions and avoid leading questions. Our interviewees were interviewed on a voluntary basis, and confidentiality was emphasised to encourage them to respond to the interview questions in the most truthful way. Secondly, during the data analysis, we adopted the following procedures consisting of two steps. In the beginning, the researchers who were the interviewers of one interview session paired with another researcher and both performed data coding for the whole transcript obtained. After that, all the researchers involved in this study were divided into three groups, with each being responsible for each research question stated in~\ref{sec:research_method:RQs} and having at least three members. All group members responsible for one research question validated the coded data related to their section in all the transcripts. At this stage, some re-codings happened in collaboration with original coders to extract more details from the data.

\textbf{External validity.} This is concerned with the limitations of how much this study can generalise conclusions. There are a few limitations associated with this study. First, although we achieved quite a spread of geography (mainly in Europe), company sizes, and business domains, they are not representative of the entire European IT economy. Had we interviewed other organisations, it is likely that the results would have differed to some extent. Although we asked the companies about their interest in sustainability, it is hard to find a common pattern and reasons for the variation of sustainability among organizations, we cannot establish a connection between particular types of companies to what kind of interest in sustainability they have. 

Second, our results can not only be subject to the inherent limitations of the codebook and the coding process but also to the biases of those interviewed and to the ontological uncertainty of the future. In particular, the frequency a code has been mentioned, while possibly representative of the perceived relative relevance within the industry nowadays, may not represent the true importance of each topic, which might only become apparent in the future. 

\textbf{Reliability.} This aspect concerns to what extent the data and the analysis depend on specific researchers. Since the researchers of this study were responsible for conducting 1-3  interviews/focus group interviews, we prepared a presentation containing all the interview questions and showed it during the meetings with our interviewees to ensure all the discussions flowed consistently. In addition, we supply the codebook as supplementary materials for validation purposes, which are helpful for replication.

\section{Related Work}\label{sec:relatedWork}
A number of studies have investigated how software engineering professionals understand sustainability. For example, Groher and Weinrich~\cite{groher2017interview} report on a qualitative interview study with ten interviews in nine organisations in Austria. They aimed to understand how practitioners understood sustainability and its importance, the factors influencing sustainability in software development, sustainability-related deficiencies in their projects, and how they improve sustainability in such projects. The results show that while practitioners find the topic of sustainability important, they seem to have a narrow view of sustainability, focusing mainly on technical attributes such as maintainability and extensibility. In addition, practitioners were concerned with organisational and economic issues, but the environmental dimension was not addressed. 

Our study differs from this one in several aspects. First, we interviewed  28 organisations spread across 9 countries, instead of just one, which potentially provides a broader and less culturally biased view of sustainability in the ICT industry. Most importantly, our interviewees had different profiles. While  Groher and Weinrich interviewed technical leaders of ICT projects, we talked to senior management and sustainability experts within the company. That can be observed in the different perceptions of sustainability. In  Groher and Weinrich's work, most interviewees related sustainability to
maintainability, extensibility, product development, and long-lived systems~\cite{groher2017interview}, while in our study, sustainability was more broadly understood, with the different dimensions mentioned. A remarkable difference is that the environment was very rarely mentioned in the previous study, while it was one of the main concerns shown in ours. However, both studies coincide in that the economic benefit is the greatest motivation for these companies. Both studies also looked into difficulties or deficiencies in sustainability. In  Groher and Weinrich's, participants mainly pointed to a ``lack of effective means for communication and knowledge sharing`` and suggested strategies such as ''knowledge distribution, avoiding specialists, and building teams that can work across products". Our study coincided with the lack of understanding of sustainability concepts and goals, yet it highlighted the trade-off between short-term financial profitability and long-term sustainability goals as a major difficulty. Our study also pointed to external difficulties, such as economic barriers and inadequate policies, which were not mentioned in the previous study.

De Souza et al.~\cite{de2014defining} discuss software sustainability as perceived by nine software developers from a university in the UK, and suggest a set of recommendations on how to improve the sustainability of software. They used short semi-structured interviews, each lasting an average of about 10 minutes. The main result is the distinction between ``Intrinsic Sustainability'', referring to intrinsic characteristics software should have (e.g., be documented, be tested, or be modular), and ``Extrinsic Sustainability'', referring to the environment in which the software is developed or used (e.g., be open, be actively maintained, or be infrastructure-independent). To address this, the authors proposed a set of recommendations as good practices for software development and maintenance that directly emerge from the characteristics interviewees associated with `intrinsic' or `extrinsic' sustainability but remain exclusively in the realm of technical sustainability. Our study differs from De Souza et al.'s~\cite{de2014defining} significantly. They interviewed software developers within a single academic organization. That meant that their participants were mostly experienced with research projects, which are of a very different nature from those of our study. Unsurprisingly, participants' views in that study were very much more related to technical sustainability, than ours. Interestingly, their questions were open and neutral, not really biasing the answers to such limited views. Finally, the coverage of the research questions as well as the depth of the interviews was very different, with ours specifically asking about goals, barriers and skills to sustainability and theirs on the relation of sustainability with software systems.  This difference in the depth can also be seen in the lengths of the interviews, which typically lasted around 10 minutes in the previous study and 1-2 hours in ours.

Karita et al.~\cite{karita2021software} report on a study performed with ninety-nine companies from the software industry in Brazil to investigate their awareness in four sustainability dimensions (environmental, economic, social and technical). The results indicate that sustainability in the context of Software Engineering is a new subject for practitioners, that they find the topic relevant, and that sustainability should be treated as a quality attribute. In contrast, this study aims further and, more concretely, at companies with such awareness to retrieve their actual interests, difficulties and achievements, and skills they have in-house and those that they miss.

In Chitchyan et. al.~\cite{chitchyan2016}, thirteen requirements engineers from eight different countries (Austria, Brazil, Germany, Spain, Switzerland, Turkey, the UK, and the USA) have been interviewed. The study investigated the perception of engineering practitioners towards sustainability as well as obstacles and mitigation strategies regarding the application of sustainable design principles in their engineering work. The study shows that on an individual level, perceptions of sustainability tend to be narrow, organisations are not aware of the potential achievements and benefits coming along with sustainable design and the standards and norms in Software Engineering are not conducive to sustainability. In contrast to our study, the work focuses on what hampers the adoption of sustainable design principles based on~\cite{becker2015sustainability} daily work practices, and not on the broader questions of industry sustainability-related interests and needs, their planned achievements, and the thus-required skills.

Other published work is more loosely related to ours, with the following worth highlighting. Betz, Lammert, and Porras~\cite{betz2022software} investigated the role of perception of software engineers; more specifically it investigates the self-attribution of software engineers and whether they implement sustainability issues in their daily work. Their results suggest that software engineers perceive that they are insufficiently involved in the design process and that they do not sufficiently take on responsibility for the software and its sustainability impacts.  The authors observed an evolution in terms of communication with interdisciplinary experts, yet their software engineers still see themselves as a ``purely executive force"~\cite{betz2022software}, who shy away from responsibility regarding sustainability. This perception varies greatly from the ones in our study, which does recognize the need for sustainability skills and competencies for IT staff. Additionally, a domain-specific study conducted by Kasurinen et al.~\cite{kasurinen2017concerns}, investigated - among other points such as the development processes used — the extent to which game developers are concerned about sustainability issues and Green IT. The results show that their interviewed gaming companies were more unstructured than general software development ones, not really incorporating sustainability in their daily work practices. Yet, our studies coincide with regard to the lack of a broader understanding of sustainability by IT professionals.

In the related field of Information Systems (IS), Cooper and Molla~\cite{cooper2017information} investigated the notion of the ``absorptive capacity'' of IS to enable environmental sustainability and how organisations can enable IS changes to address environmental issues. They conducted a survey with 148 IS senior managers and provided different taxonomies to acquire knowledge about sustainable IS and to what extent IS sustainable technologies are assimilated by organisations. The role of ``absorptive capacity'' is also discussed in~\cite{dzhengiz2020competences} where the authors provide a systematic literature review on competencies for environmental sustainability and managerial skills required for organisations to transform knowledge into environmental capabilities. The work suggests a connection between environmental competencies and capabilities, and they provide a taxonomy between management and environmental competencies.

\section{Conclusions and future work}\label{sec:conclusion}
Our study has uncovered how sustainability is viewed and practised in 28 organisations from nine countries. The findings of this work include (i) how sustainability is of interest to these organisations, (ii) sustainability goals they want to achieve, (iii) difficulties they have encountered so far, (iv) skills and competencies needed to achieve the established goals, and (v) current practices to fill the perceived sustainability skill gap. 
Identifying those current practices and especially the gaps, gives us an indication of possible improvements to current university education programs with respect to sustainability for IT and related fields. This study represents the first step to improving the computing curricula to better meet the demands of industry. To accelerate that process, we have proposed essential topics relevant to sustainability in IT that should be taken into account when developing the curricula, based on our years of experience in teaching courses in sustainability for IT students.

We also highlighted several open research opportunities, directions, and challenges: 
First, while significant, the organisations we interviewed provide only a partial geographical perspective; our analysis should be performed globally.
We thus plan to survey on a global scale to obtain a more comprehensive picture and to be able to conduct quantitative analyses, for example, regarding the variation of sustainability among organisations.
As it is easier to include more companies in a survey than in an interview series, this will also allow the mapping of individual business domains and different sustainability perspectives. 
Finally, the ultimate goal should be the rapid development of software engineering and computer science curricula which include sustainability concepts at their very core. 
We presented first ideas in Section~\ref{sec:discussion}.
These curricula should certainly be holistic and not aim exclusively at the skills needed by industry. However, given the growing sustainability interest of the industry and its immense transformational power, the curricula should definitely take its sustainability needs into consideration.

\ifCLASSOPTIONcompsoc
  \section*{Acknowledgments}
\else
  \section*{Acknowledgment}
\fi

The authors would like to thank all the interviewees who took part in the study.

\ifCLASSOPTIONcaptionsoff
  \newpage
\fi



%
\balance
{\footnotesize
\bibliographystyle{IEEEtran}
}
\bibliography{main_v3}

\end{document}